\documentclass[twocolumn]{article}
\usepackage[left=2.50cm, right=2.50cm, top=2.50cm, bottom=2.50cm]{geometry}
\usepackage{helvet}
\usepackage[english]{babel}
\usepackage{booktabs}
\usepackage{subfigure}
\usepackage{caption}
\usepackage{algorithm}
\usepackage{algorithmic}
\usepackage{graphicx}
\usepackage{listings}
\usepackage{diagbox}
\usepackage{amsmath}
\numberwithin{equation}{section}
\usepackage{amsfonts}
\usepackage{amsthm}
\usepackage{esint}
\usepackage{color}
\usepackage{fancyhdr}
\usepackage{hyperref}
\hypersetup{
    colorlinks=true,
    linkcolor=blue,
    citecolor=blue,
    filecolor=blue,
    urlcolor=blue,
}
\usepackage[backend=biber,style=numeric-comp,sorting=none,isbn=false,url=false]{biblatex}
\renewbibmacro{in:}{}
\usepackage{authblk}

\title{Coarse-grained binning in Drell-Yan transverse momentum spectra}
\author[1]{Wenxiao Zhan}
\author[1]{Siqi Yang \thanks{Email: \href{mailto:yangsq@ustc.edu.cn}{yangsq@ustc.edu.cn}}}
\author[1]{Minghui Liu}
\author[2,3]{Francesco Hautmann}
\author[1]{Liang Han}
\affil[1]{University of Science and Technology of China, Hefei, China}
\affil[2]{University of Oxford, Oxford, UK}
\affil[3]{University of Antwerp, Antwerp, Belgium}
\date{}

\begin{document}
\maketitle

\begin{abstract}
     We report a study of the determination of the intrinsic transverse momentum of partons, the intrinsic $k_T$, from the dilepton transverse momentum $p_T$ in Drell-Yan (DY)
    production at hadron colliders. The result shows that a good sensitivity to
    the intrinsic  $k_T$ distribution
     is achieved by measuring relative ratios between the cross sections of
     suitably defined low-$p_T$ and high-$p_T$  regions. The study is performed through both a pseudo-data test and an extraction from measurements of the DY process by the CMS collaboration. Since the methodology does not rely on any dedicated partition of bins, this $p_T$-ratio observable requires less special treatment in very low $p_T$ regions, and propagates lower systematic uncertainties induced from unfolding or momentum migration, in contrast with previous proposals of using a fine-binning measurement of the differential cross section.
\end{abstract}

\section{Introduction}

Measurements of transverse momentum spectra of
electroweak bosons via
Drell-Yan (DY) lepton-pair production~\cite{Drell:1970wh}
are at the core of many aspects of the physics program at the
Large Hadron Collider (LHC),  ranging
 from precision determinations of Standard Model parameters
in the strong and electroweak sectors,  to
non-perturbative features of  hadron structure  and
transverse momentum dependent (TMD)
parton distribution functions~\cite{Angeles-Martinez:2015sea}.

 Recent  applications of
DY transverse momentum
 measurements
include the determination of the strong coupling
from perturbative Quantum Chromodynamics (QCD) predictions matched with
resummation~\cite{Camarda:2022qdg,ATLAS:2023lhg};
the extraction   of  TMD
 parton distributions both
in
analytic-resummation~\cite{Bacchetta:2022awv,Bacchetta:2024qre,Bury:2022czx,Moos:2023yfa} and parton-branching~\cite{Bubanja:2023nrd}
approaches;
the modeling of TMD contributions in
soft-collinear effective theory~\cite{Billis:2024dqq};
the determination of intrinsic-$k_T$ parameters in the  tuning of
Monte Carlo event generators~\cite{CMS:2024eprint}.

A common thread in these applications is that high-precision
measurements of the low transverse-momentum
region
($\Lambda_{\rm{QCD}}
{\raisebox{-.6ex}{\rlap{$\,\sim\,$}} \raisebox{.4ex}{$\,<\,$}} p_T(ll) \ll m(ll)$,
where $p_T$ and $m$ are the dilepton's transverse momentum and
invariant mass)
with fine binning in $p_T$ will improve our ability
to unravel QCD dynamics,  involving multiple soft-gluon emissions
as well as the non-perturbative intrinsic transverse motion of partons.
See for instance the role of  fine-binned $p_T (ll)$ measurements
in the context of analytic-resummation studies~\cite{Hautmann:2020cyp}
and parton-branching studies~\cite{BermudezMartinez:2019anj}.

However, recent measurements of the DY process by the
ATLAS~\cite{RN60} and CMS~\cite{RN66} collaborations indicate
that such fine-binned treatments require  an extremely delicate control
of systematic uncertainties.
Experimentally,  the determination of
the low-$p_T(ll)$ fine-binned structure
is
challenging.
Under these circumstances, it
becomes important to assess the capabilities of methodologies
which do not require the fine-binned measurement of
$p_T(ll)$ distributions, and rather rely on coarse-grained binning.

In this paper we explore one such methodology, by studying
the  intrinsic $k_T$ determination in TMD parton distributions
from the measurement of $ pp \to Z / \gamma \to l^+ l^- $
in the region $0 < p_T(ll) < p_{T , {\rm{max}}} $ (with $p_{T , {\rm {max}}} $ small
compared to   $ m(ll)$)    based on coarse-grained partitions of this region.
We
subdivide the $p_T$ region into two bins with separation momentum $p_s$,
$p_T < p_s$ and $p_T > p_s$,  and
investigate the sensitivity to the intrinsic $k_T$
distribution from measuring the ratio between the
low $p_T (ll)$ cross section (predominantly sensitive to
TMD dynamics and resummed soft-parton radiation) and
the high $p_T(ll) $ cross section
 (predominantly sensitive to  fixed-order hard-parton radiation)
as a function of varying $p_s$. We
study the role of
bin-to-bin migration effects on the $p_T$ ratio.
Although the high $p_T(ll)$ cross section  is
not sensitive to intrinsic $k_T$, it  acts as a reference
in the relative ratio between the strength of the
intrinsic $k_T$ and fixed-order contribution.
We find that
the intrinsic $k_T$ can be determined by measuring its overall strength
through the $p_T$-ratio instead of measuring the low $p_T(ll)$ differential cross section.
This conclusion is relevant, from a practical viewpoint,  in order to achieve
a good determination of intrinsic $k_T$
 from analyses of experimental data with
reduced systematic uncertainties.
We
illustrate this methodology
by presenting an extraction of
intrinsic $k_T$ from
DY  $p_T(ll)$ measurements by the CMS
collaboration~\cite{CMS:2022ubq}.

To perform this study, we use the parton branching (PB)
approach~\cite{Hautmann:2017xtx,Hautmann:2017fcj} to
TMD evolution.  This approach provides the basis
for including TMD distributions in parton shower Monte Carlo
calculations, and has been shown to
give a consistent description of
DY $p_T$ spectra~\cite{Bubanja:2023nrd}
as well as of the multiple-jet structure associated
with DY production~\cite{BermudezMartinez:2022bpj}.
At inclusive level, it can be related to the evolution of collinear
parton distribution functions (PDF), and   correctly describes
deep-inelastic scattering
structure functions~\cite{BermudezMartinez:2018fsv}.
Given the wide applicability of the PB TMD approach, this
provides a useful framework to assess the performance of
the $p_T$-ratio methodology.

The paper is organized as follows. In Sect.~\ref{sec:pb}, we briefly review the
PB TMD approach. In Sect.~\ref{sec:ss}, we propose the $p_T$-ratio as a new
observable and methodology for studies of TMD dynamics. The sensitivity
of the $p_T$-ratio to the intrinsic $k_T$ distribution is investigated with a
pseudo-data sample, and is compared to that from the fine-binning structure of
the low-$p_T(ll)$ differential cross section. In Sect.~\ref{sec:meas}, we apply
the methodology to the $p_T(ll)$ distribution measured by the CMS
collaboration~\cite{CMS:2022ubq} at 13 TeV across a wide range in
DY mass from 50 GeV to 1 TeV. We give
conclusions in Sect.~\ref{sec:conclusion}.

\section{PB TMD method}\label{sec:pb}

In this section we briefly describe the main features of the PB
approach~\cite{Hautmann:2017xtx,Hautmann:2017fcj,BermudezMartinez:2018fsv}
which will be used for the analysis in this work.
We first recall the PB evolution equations; then we discuss the
  intrinsic-$k_T$ distribution, the  soft-gluon resolution scale,
  the treatment of the strong coupling, and the
  application of the method  to the computation of DY $p_T$ distributions.

We consider
 the TMD parton distribution of  parton flavor $a$,
 $A_a  ( x, {\bf k}, \mu^2 )$,
as a function  of  the
longitudinal momentum  fraction $x$,
 the transverse momentum ${\bf k}$ and  the evolution scale $\mu$.
 According to Refs.~\cite{Hautmann:2017fcj,BermudezMartinez:2018fsv},
the  distributions $A_a  ( x, {\bf k}, \mu^2 )$
fulfill evolution equations of the   form
\begin{eqnarray}
\label{eq:2bin_tmdevol}
&&  {A}_a ( x, {\bf k}, \mu^2 ) =
 \Delta_a  (\mu^2, \mu_0^{2} )
 {A}_a ( x, {\bf k}, \mu_0^2 )
\nonumber  \\
 &+&
 \sum_b\int \frac{\textrm{d}^2{\boldsymbol \mu}^{\prime}}{\pi {\mu}^{\prime 2}}
   \int  \textrm{d}z  \
   {\cal E}_{a b}  [ \Delta ;  P^{(R)};  \Theta
]
\nonumber   \\
 & \times &
{A}_b ( x / z ,  {\bf k} + (1-z){\boldsymbol \mu}^\prime, \mu^{\prime 2}) \; ,
\end{eqnarray}
where  $ \Delta_a (\mu^2, \mu_0^{2} ) $
 is the Sudakov form factor
and
 ${\cal E}_{a b} $ are the evolution kernels,
 which are specified  in Ref.~\cite{Hautmann:2017fcj}
 as  functionals of the Sudakov form factors $\Delta_a$, of the
 real-emission splitting functions   $ P_{ab}^{R}$, and
 of phase space constraints collectively denoted by
 $\Theta$ in Eq.~(\ref{eq:2bin_tmdevol}).
 The functions that appear in the evolution kernels are perturbatively
 computable as power series expansions in the strong coupling $\alpha_s$. The
 explicit expressions of these expansions for all flavor channels
 are given to two-loop order in Ref.~\cite{Hautmann:2017fcj}.

 The evolution in Eq.~(\ref{eq:2bin_tmdevol})
is expressed in terms of two branching variables,
$z$ and ${\boldsymbol \mu}^\prime$: $z$ is the longitudinal momentum transfer
at the  branching,
controlling the  rapidity of
the parton emitted along the  branching chain;
 $ \mu^\prime = \sqrt{ {\boldsymbol \mu}^{\prime 2}}$ is the mass
scale at which the branching occurs, and is related
  to the branching's kinematic variables according to the
  ordering condition.
  It is worth noting that the double evolution in
rapidity and mass  for TMD distributions can also be
formulated in a CSS~\cite{Collins:1984kg,Collins:2011zzd}, rather
than PB, approach ---   see
e.g.~\cite{Scimemi:2018xaf,Hautmann:2020cyp,Hautmann:2021ovt}.
A study of the relationship of the PB Sudakov evolution kernel
with the CSS formalism is performed in~\cite{Martinez:2024twn,Lelek:2024kax}.
  Eq.~(\ref{eq:2bin_tmdevol}) is obtained
  for the case of
   angular
  ordering~\cite{Marchesini:1987cf,Catani:1990rr,Hautmann:2019biw},
  which gives $\mu^\prime  = q_{\perp  }  / (1-z)$,  where $q_{\perp  }$
  is the emitted parton's transverse momentum.
The angular-ordered branching is motivated by the treatment of the endpoint
region in the TMD case~\cite{Hautmann:2017fcj,Hautmann:2007uw}.

    The initial
evolution scale in Eq.~(\ref{eq:2bin_tmdevol})
is denoted by $\mu_0$
($ \mu_0 >   \Lambda_{\rm{QCD}}  $), and is  usually
taken to be of order 1 GeV.
  The   distribution
  $ {A}_a  ( x, {\bf k}, \mu_0^2 )$
  at scale $\mu_0$ in the first term on the right hand side of
Eq.~(\ref{eq:2bin_tmdevol})   provides the intrinsic   $k_T$  distribution.
   This is  a nonperturbative
boundary condition to the evolution equation, and may
 be determined from comparisons of
theory predictions to experimental data.
In the calculations of this paper, for simplicity
we will parameterize  $  { {\cal A}}_a(x,{\bf k},\mu^2_0)   $,
following  previous applications of the
PB TMD method~\cite{BermudezMartinez:2018fsv,Bubanja:2023nrd},
as
\begin{eqnarray}
\label{TMD_A0_2bin}
 {\cal A}_{0,b} (x, k_T^2,\mu_0^2)  & = & f_{0,b} (x,\mu_0^2)
\\
&  \times & \exp\left(-| k_T^2 | / 2 \sigma^2\right) / ( 2 \pi \sigma^2) \; ,
\nonumber
\end{eqnarray}
with the width of the Gaussian
distribution given by
$ \sigma  =  q_s / \sqrt{2} $,
independent of parton flavor  and $x$, where
$q_s$ is the intrinsic-$k_T$ parameter.

The evolution kernels $ {\cal E}_{a b} $
 and   Sudakov form factors $\Delta_a$  in Eq.~(\ref{eq:2bin_tmdevol})
 contain kinematic
 constraints  which embody the phase space of the branchings
 along the parton cascade. These can be described, using the
 ``unitarity'' picture of QCD evolution~\cite{Webber:1986mc},
by separating resolvable and non-resolvable branchings
 in terms of a  resolution scale to classify
 soft-gluon emissions~\cite{Hautmann:2017xtx}.
 (See e.g.~\cite{Dooling:2012uw} for a study of the interplay
 of resolution scale with transverse momentum recoils
 in parton showers.)
 Applications
of  the PB TMD method can be carried out   either by taking
the soft-gluon resolution  scale
  to be a fixed constant value close to the
 kinematic limit  $z =1$
 or by allowing for a running, ${ \mu}^\prime$-dependent
 resolution (see, e.g., discussions
 in~\cite{Hautmann:2019biw,Barzani:2022msy,Bubanja:2023nrd}).
In the computations of this paper, we will take  fixed resolution scale.
This is the same as what is done in the
studies of Refs.~\cite{BermudezMartinez:2018fsv,BermudezMartinez:2019anj,Bubanja:2023nrd}, which
we will use as benchmarks.

The scale at which the strong coupling $\alpha_s$ is
to be evaluated in
Eq.~(\ref{eq:2bin_tmdevol})
is a function of the branching variable.
Two scenarios are commonly studied in PB TMD applications:
 i)   $\alpha_s =  \alpha_s ( { \mu}^{\prime 2} )$;
ii) $\alpha_s = \alpha_s ( q_\perp^2 )
=  \alpha_s ( { \mu}^{\prime 2}  (1-z)^2 )$.
Case~i)  corresponds to
DGLAP
evolution~\cite{Gribov:1972ri,Altarelli:1977zs,Dokshitzer:1977sg} , while
case~ii) corresponds to angular ordering,
e.g.~CMW evolution~\cite{Marchesini:1987cf,Catani:1990rr}.
In Ref.~\cite{BermudezMartinez:2018fsv}, fits to
precision deep inelastic scattering
HERA data~\cite{Abramowicz:2015mha}
are performed for both  scenarios i)  and   ii),
using the   fitting platform
  \verb+xFitter+~\cite{xFitter:2022zjb,Alekhin:2014irh}.
It is found that  fits with good $\chi^2$
values can be achieved
in either case. Correspondingly, PB-NLO-2018 Set1
(with  the DGLAP-type
$ \alpha_s ({\bf q}^{\prime 2}) $)
and PB-NLO-2018 Set2 (with the angular-ordered CMW-type
$ \alpha_s  (q_T^2)$) are obtained.

On the other hand,   it is found that
 PB-NLO-2018 Set2 provides a much better
 description, compared to PB-NLO-2018 Set1, of
 measured $Z / \gamma$  $p_T$  spectra
 at the LHC~\cite{BermudezMartinez:2019anj} and  in
low-energy experiments~\cite{BermudezMartinez:2020tys}, and
of  di-jet azimuthal correlations near the back-to-back
region at the LHC~\cite{Abdulhamid:2021xtt,BermudezMartinez:2022tql}.
Further, it is shown that a good description is also
obtained for DY + jets final-state distributions~\cite{BermudezMartinez:2021lxz,BermudezMartinez:2021zlg,BermudezMartinez:2022bpj,Yang:2022qgk}.
 We will therefore base the
 analysis of this work on PB-NLO-2018 Set2.
 As in~\cite{BermudezMartinez:2018fsv,Bubanja:2023nrd},
the strong coupling is modeled according to
a ``pre-confinement'' picture~\cite{Amati:1980ch,Bassetto:1983mvz}
as $\alpha_s = \alpha_s(\max(q^2_{c},{\bf q}_{\perp}^2))$,
where $q_c $ is a semi-hard
scale, taken to be $q_c = 1$ GeV.
 The PB TMD sets are available from the
 {\sc TMDlib} library~\cite{Abdulov:2021ivr,Hautmann:2014kza}.

The calculation of DY production cross sections
in the PB TMD method  proceeds
as described in
Ref.~\cite{BermudezMartinez:2019anj}.
NLO hard-scattering matrix elements are
obtained from the
{\scshape MadGraph5\_aMC@NLO}~\cite{Alwall:2014hca}
(hereafter, MCatNLO)  event generator
and matched with TMD parton distributions and showers
obtained from PB evolution~\cite{BermudezMartinez:2018fsv,Hautmann:2017fcj,Hautmann:2017xtx} and implemented in the {\sc Cascade} Monte Carlo
event generator~\cite{CASCADE:2021bxe,CASCADE:2010clj},
using the subtractive matching procedure proposed
in~\cite{BermudezMartinez:2019anj} and further
analyzed in~\cite{Yang:2022qgk}. In particular, the
 {\sc Herwig}6~\cite{Corcella:2002jc,GennaroCorcella_2001} subtraction
 terms are used in MCatNLO,
 as they are based on the same angular ordering conditions
 as the PB TMD distributions~\cite{Yang:2022qgk}.
Final state parton showers
are generated from  {\sc Pythia}6.428~\cite{pythia}, including
photon radiation.

Ref.~\cite{Bubanja:2023nrd} uses the approach described above
to compute  theoretical results
for DY transverse momentum
distributions, and
to  make a determination
 of the  intrinsic-$k_T$ parameter
$q_s$ in Eq.~(\ref{TMD_A0_2bin})
by performing  fits  of the results to DY experimental
measurements of DY $p_T(ll)$~\cite{CMS:2022ubq,ATLAS:2015iiu,LHCb:2021huf,D0:1999jba,CDF:1999bpw,CDF:2012brb,PHENIX:2018dwt,CMS:2021ynu,Moreno:1990sf}.
The study~\cite{Bubanja:2023nrd} includes a detailed treatment of
 statistical, correlated and uncorrelated
uncertainties.  The
CMS data~\cite{CMS:2022ubq} at center of mass energy
$\sqrt{s} = $ 13 TeV cover DY invariant masses from 50 GeV to 1 TeV. The other
data cover energies from 13 TeV down to 38 GeV. With
the fitted $q_s$ values~\cite{Bubanja:2023nrd},
PB TMD distributions are thus obtained.
Further discussions of the PB TMD distributions~\cite{Bubanja:2023nrd}
are given in~\cite{Jung:2024eam,Bubanja:2024puv,Monfared:2024vgc,Raicevic:2024obe}.

In the next section we will use
Monte Carlo predictions
from the PB TMD method
to explore the feasibility of extracting
intrinsic-$k_T$ distributions from experimental
measurements with coarse-grained binning in
the low $p_T(ll)$  region.

\section{Ratio of low and high $p_T$ bins}\label{sec:ss}

In this section we present  the $p_T$-ratio
observable and methodology for studies of TMD dynamics.
We will focus on the process $ pp \to Z / \gamma \to l^+ l^- $
in the transverse momentum
region $0 < p_T(ll) < p_{T , {\rm{max}}} $, where
 $p_{T , {\rm {max}}} $ is taken to be small compared to
 the invariant mass $ m(ll)$.

\begin{figure}[h]
    \centering
    \includegraphics[width=\linewidth]{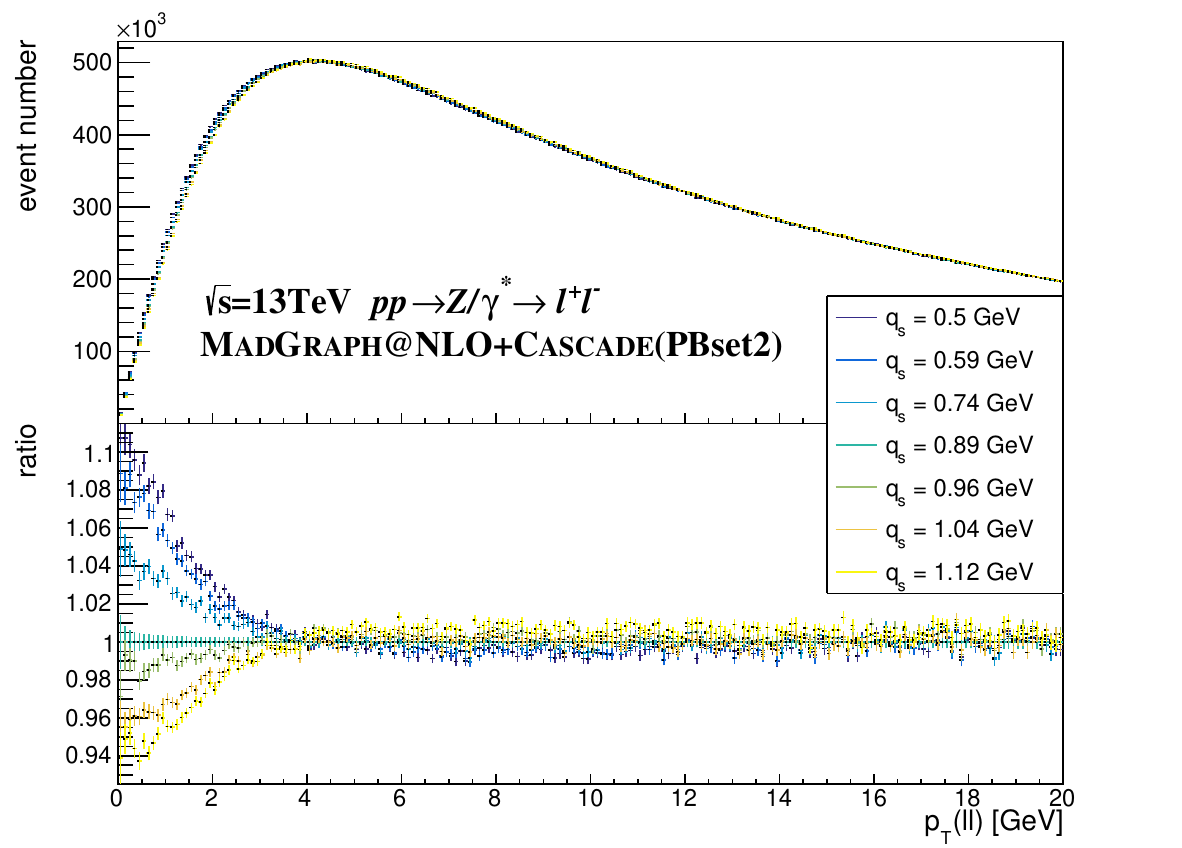}
    \caption{Transverse momentum distributions from
    the template samples
    with different intrinsic $k_T$ Gaussian width $q_s$ in full phase space. The ratio plot
    at the bottom is made by taking the ratio to the distribution with $q_s$ = 0.89.}
    \label{fig:pt}
\end{figure}

We construct
PB TMD predictions for the DY $p_T(ll)$ distribution,
as described in the previous section,
 from MCatNLO + {\sc Cascade} Monte Carlo,
 with initial-scale TMD distributions given in
 Eq.~(\ref{TMD_A0_2bin}).
  To explore the  intrinsic-$k_T$   parameter space,
we generate 7 template samples with
different $q_s$ values using the TMD grid files available
on the  {\sc TMDlib}~\cite{Abdulov:2021ivr} website: these are
$q_s$ = 0.5 GeV (using the PB TMD set
\textit{PB-NLO-HERAI+II-2018-set2}~\cite{BermudezMartinez:2018fsv}), $q_s$ = 0.59, 0.74, 0.89 GeV (using the set  \textit{PB-NLO-HERAI+II-2023-set2-qs=0.74}~\cite{Bubanja:2023nrd}), and $q_s$ = 0.96, 1.04, 1.12 GeV (using the set \textit{PB-NLO-HERAI+II-2023-set2-qs=1.04}~\cite{Bubanja:2023nrd}).
We study the influence of the intrinsic-$k_T$ width by examining
the templates firstly in the full phase space (later on we will consider an experimental fiducial phase space). Results
are shown  in Fig.~\ref{fig:pt} over the range $ p_T(ll) <  20$ GeV.

\begin{figure}[h]
    \centering
    \includegraphics[width=\linewidth]{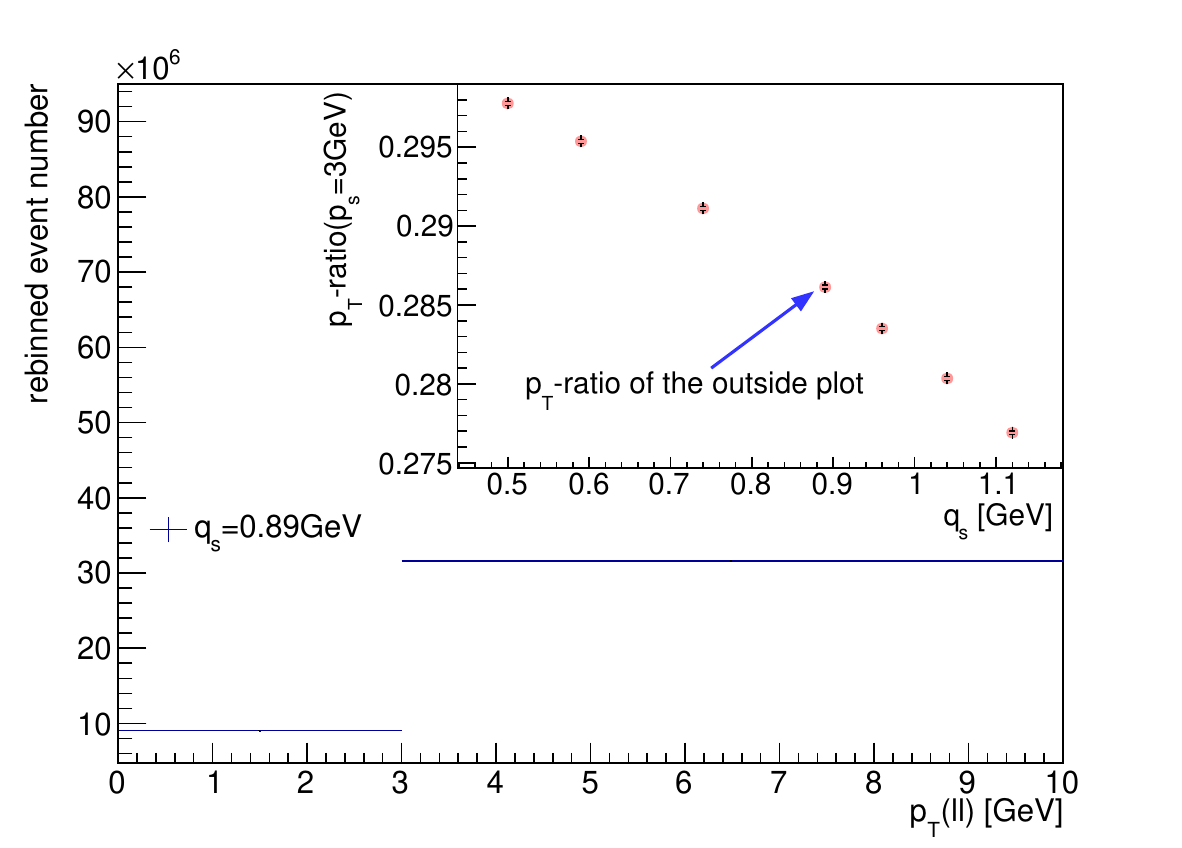}
    \caption{The 2-bin distribution of lower and higher $p_T$ (outer) and the corresponding $p_T$-ratio versus $q_s$ (inner).}
    \label{fig:show_ratio}
\end{figure}

The $p_T(ll)$ shape in the rising part of the spectrum is altered by
 $q_s$, with the peak position shifting to higher $p_T$ when $q_s$ increases.
 However, the trend is mild, suggesting that the essential information on
 the intrinsic $k_T$ width may be extracted from a well-defined ratio between
 the low and high $p_T$ regions. To this end, we introduce a momentum
 parameter $p_s$ to separate the regions $p_T < p_s$ and $p_T > p_s$, and
 construct the 2-bin $p_T$-ratio between the lower and higher $p_T$ regions.
 The $p_T$-ratio is defined from the integral of event numbers in the
 relatively low ($p_L$) and high ($p_H$) region, i.e., from the two-binned $p_T(ll)$,
\begin{equation}\label{eq:pt-ratio}
    p_T\mathrm{-ratio} = {p_L} / {p_H} .
\end{equation}
The uncertainty of the $p_T$-ratio is obtained from  propagating  statistical
uncertainties in the low and high $p_T$ bins, and is given by
\begin{equation}
    \sigma_{p_T\mathrm{-ratio}}^2 =
    ({p_L^2\sigma_{p_H}^2 + p_H^2\sigma_{p_L}^2} )/ {p_H^4}   .
\end{equation}
The 2-bin distribution and corresponding $p_T$-ratio are plotted, for
the case of a given value  $p_s = 3 $ GeV,   in Fig.~\ref{fig:show_ratio}.
The fall-off of the $p_T$-ratio   with increasing $q_s$ indicates that
one may be able to extract TMD parameters from this ratio instead
of the complete fine-binned $p_T$ shape, which results into a significant
advantage  from the standpoint of controlling experimental systematic
uncertainties.

The separation momentum $p_s$ will play a critical role
in the $q_s$ extraction from the $p_T$-ratio. TMD sensitivity arises
primarily from the low-$p_T$ region. If $p_s$ is too low inside this region,
a proportion of soft-gluon splitting transverse momenta will be integrated
into the upper region, causing a loss of sensitivity to TMD parameter
determination. If, on the other hand, $p_s$ is higher than the peak, the
information on low $p_T$ will tend to become undetectable.

We next perform a sensitivity test on both the fine-binned $p_T$ and the $p_T$-ratio. We test the former with different bin width from 0.5 GeV to 3 GeV, to examine the potential of the fine-binned $p_T$ shape in extracting $q_s$. We test the latter, on the other hand, focusing on the sensitivity change due to different choice of $p_s$. We shift $p_s$ from 1.5 GeV to 4 GeV to find an optimal definition of the $p_T$-ratio, as well as to verify whether it has statistically the same sensitivity as the fine-binned $p_T$ shape. The $p_T(ll)$ greater than 10 GeV is expected to be ancillary in TMD performance studies. In this test, we will consider $p_{T,{\rm{max}}} = $ 10~GeV,  20~GeV, in a similar spirit to previous studies of
intrinsic-$k_T$ distributions (see e.g.~Refs.~\cite{Moos:2023yfa,Bubanja:2023nrd,Hautmann:2020cyp}).

\begin{figure}[h]
    \centering
    \includegraphics[width=\linewidth]{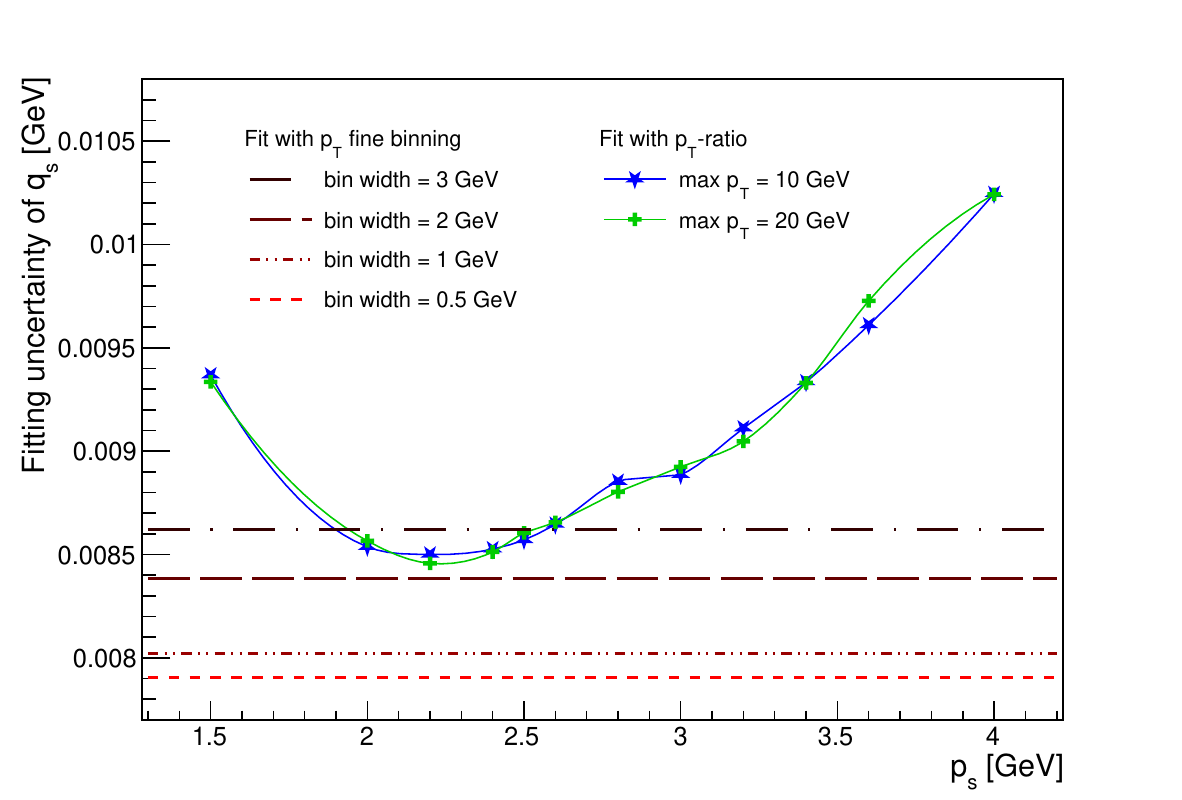}
    \caption{Uncertainties from binned $p_T$ and from $p_T$-ratio. The uncertainties from different bin widths of binned $p_T$ are drawn in straight lines, while the uncertainties given by $p_T$-ratio are marked and smoothed by curves. The abscissa of each marker is the corresponding $p_s$.}
    \label{fig:ss}
\end{figure}

In this test, we choose the sample with $q_s=0.89$ as the pseudo-data, and extract $q_s$ with the PB TMD replicas in the phase space with final-state lepton selections $p_T(l^\pm)>25$ GeV and $|\eta(l^\pm)|<2.4$, in which leptons are dressed with photons reconstructed within $\Delta R=\sqrt{(\Delta\eta)^2+(\Delta\phi)^2}<0.1$. This is close to the fiducial phase space in real experiments at ATLAS and CMS. Sensitivity is represented by the fitting uncertainty of $q_s$. Each sample in PB TMD template contains 100 million events. The fitting is performed with least square method, in which the estimator $\chi^2$ is defined as
\begin{equation}
    \chi^2=\sum_{ij}(m_i-\mu_i)C^{-1}_{ij}(m_j-\mu_j)
\end{equation}
with $m_i$ and $\mu_i$ being the observed and predicted data point, respectively, and $C_{ij}$ the covariance matrix. We consider uncorrelated statistical uncertainties from the pseudo-data and template as the only sources contributing to $C_{ij}$, making it a diagonal matrix. The fitting uncertainties in both cases are depicted in Fig.~\ref{fig:ss}. The result indicates that the $p_T$-ratio is as sensitive as the fine-binned $p_T$ shape.

\begin{figure}[h]
    \begin{subfigure}
        \centering
        \includegraphics[width=\linewidth]{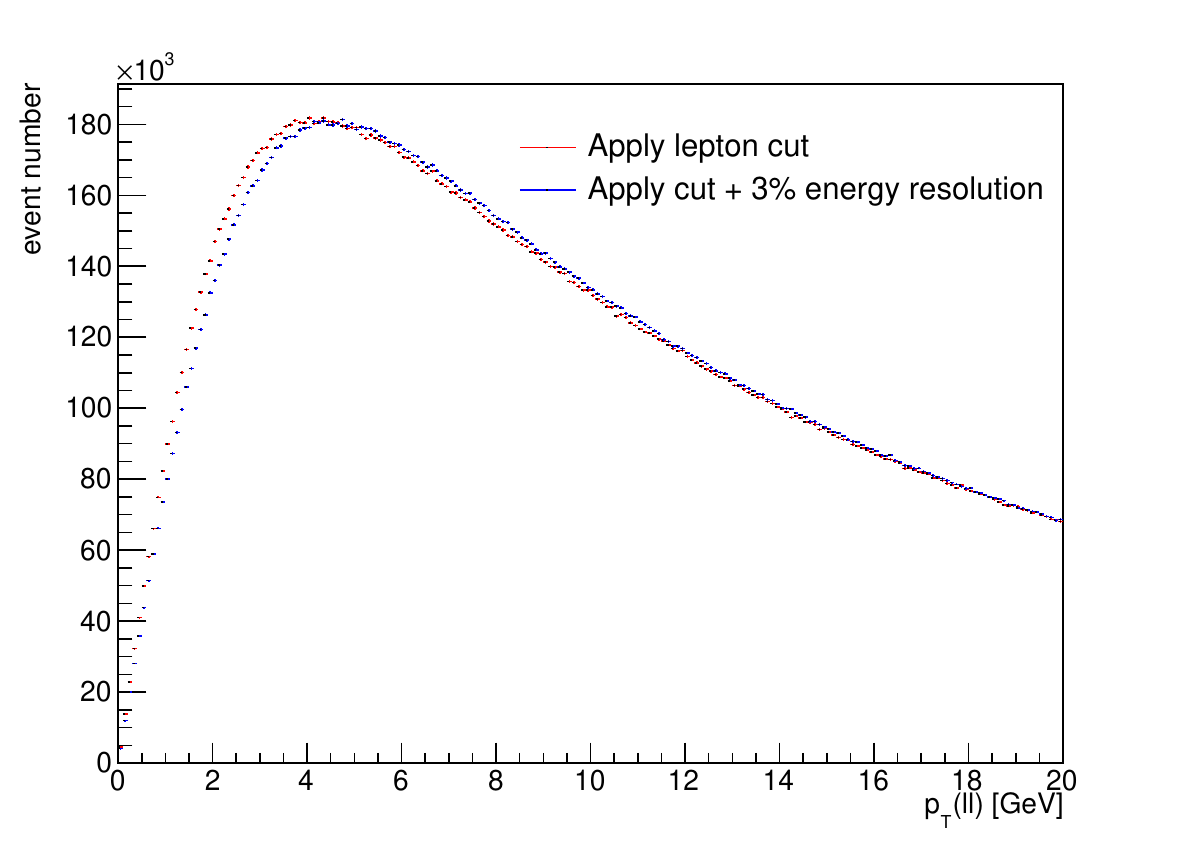}
    \end{subfigure}
    \begin{subfigure}
        \centering
        \includegraphics[width=\linewidth]{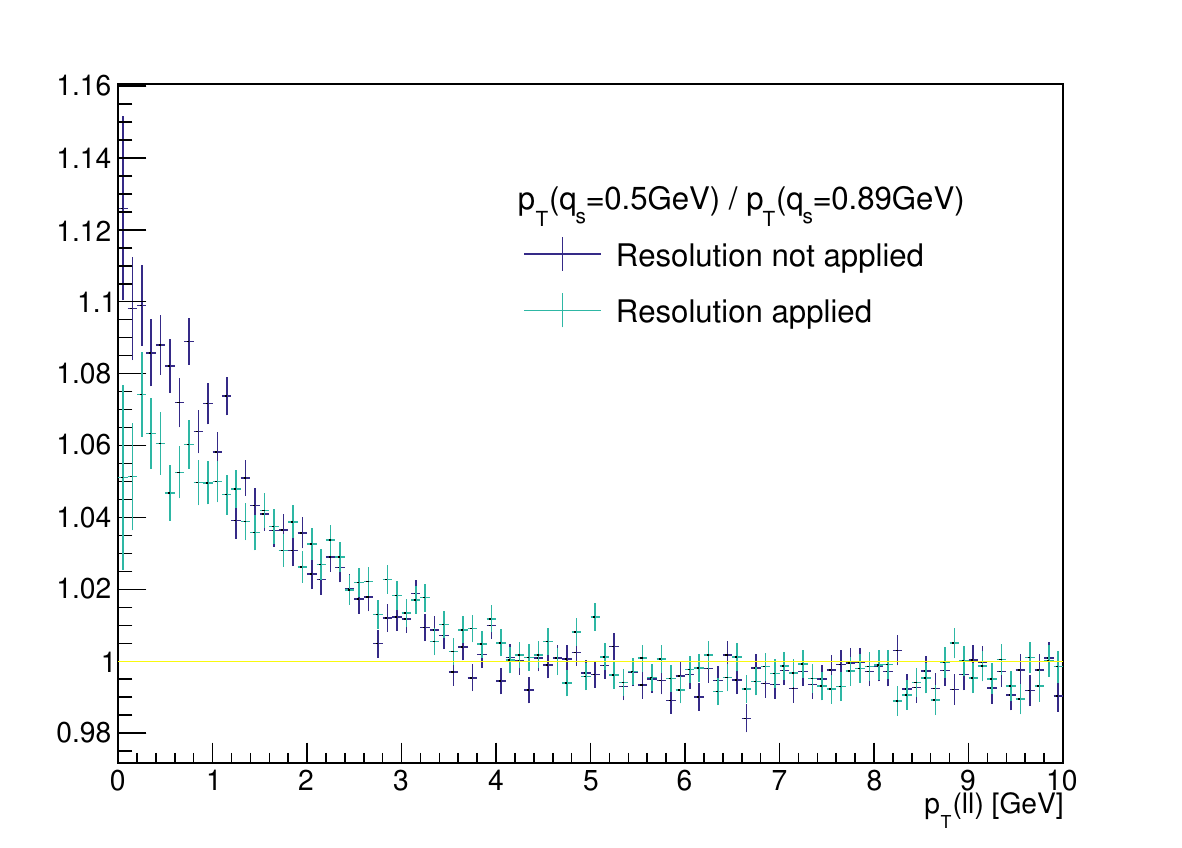}
    \end{subfigure}
    \caption{The  $p_T(ll)$ distributions when the resolution effect is taken into account. The top plot shows the difference in the $p_T(ll)$ distribution before and after this effect. The bottom plot shows the difference in the relative ratio between
    the  $q_s = 0.5$  GeV  and  $q_s = 0.89$ GeV cases.}
    \label{fig:mig}
\end{figure}

In real experiments, momentum resolution causes bin-to-bin migration. Since the migration effect is a dominant systematic uncertainty, it is essential to verify that such effect on the $p_T$-ratio is not destructive. To study this, we apply a 3\% resolution on the four-momentum of the dressed leptons as
\begin{equation}
    p^{\mu}_{reco}(l^\pm)=p^{\mu}(l^\pm)\times(1+g)
\end{equation}
where $g$ follows a Gaussian distribution $\mathcal G(0,0.03)$.
The migration effect is illustrated in Fig.~\ref{fig:mig}. It moves the peak position to a higher $p_T$ value, which is a  similar trend as the $q_s$ effect. Also, owing to this the optimal separation choice $p_s$ is different from the non-migration scenario. If $p_s$ is defined correctly in this case, we expect the behavior to be the same as it is at truth level. We also note that in the bottom plot in
Fig.~\ref{fig:mig} the relative ratio for $p_T(ll)<1$ GeV is smaller at the reco-level. This should result in a mild decrease in sensitivity.

\begin{figure}[h]
    \centering
    \includegraphics[width=\linewidth]{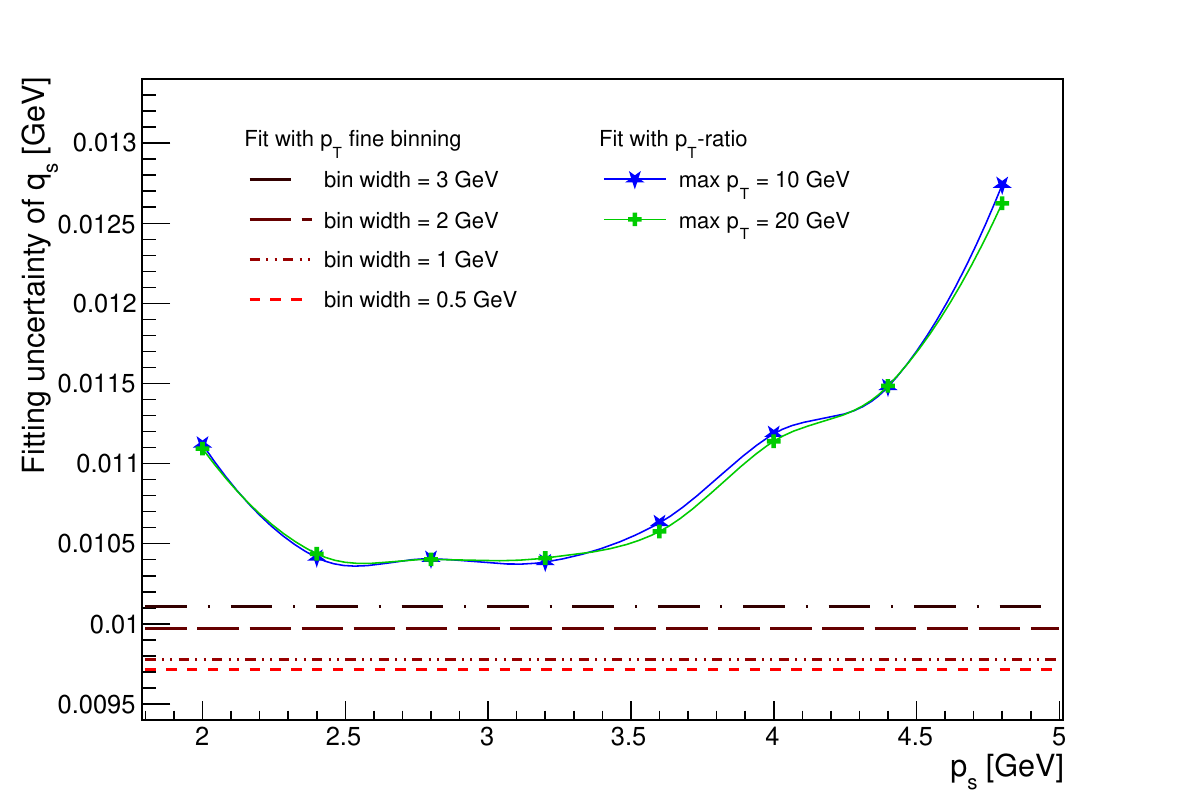}
    \caption{Uncertainties from binned $p_T$ and from $p_T$-ratio with 3\% energy resolution applied.}
    \label{fig:ssreco}
\end{figure}

We now repeat the sensitivity test after the 3\% energy resolution is applied to the dressed leptons. The results are shown in
Fig.~\ref{fig:ssreco}. The uncertainties are slightly larger than those in Fig.~\ref{fig:ss}.
Also, by comparing the fitting uncertainties in the cases of
fine-binned $p_T(ll)$ and $p_T$-ratio, we see that the decrease in sensitivity due to the migration effect is similar.  We are led to conclude that the $p_T$-ratio still has a good sensitivity to the intrinsic $k_T$, so that this methodology can be applied directly in the experiments for TMD performance studies.

\section{Extraction of $q_s$ from $p_T$-ratio}\label{sec:meas}

Having performed a test with pseudo-data in the previous section, in this section
we move on to consider a real experimental data analysis.

DY $p_T(ll)$ spectra are measured at the LHC with bin widths
equal to or greater than 1 GeV. Measuring the  fine-binned $p_T(ll)$
 structure, on the other hand, is challenging. Sizeable systematic uncertainties have been reported in recent DY differential cross section measurements~\cite{RN60, RN66},  with luminosity of approximately 36 $fb^{-1}$ data collected in 2016, especially for the $p_T(ll)<2$ GeV bins. This is due to the uncertainties mostly induced from lepton efficiency and momentum resolution.  Moreover, since the bin-to-bin correlation of the unfolding method uncertainties and momentum migration uncertainties are negative, increasing the bin numbers in such regions inevitably increases these systematic uncertainties.

Given these difficulties in achieving  finer binning in the very low $p_T(ll)$ range,
we next test the $p_T$-ratio methodology of the previous section, based on
less fine binning in the  low-$p_T$ fiducial region,  by applying it to real
experimental data.  We emphasize that the purpose of this study is not
to carry out a precision extraction of $q_s$, but to perform a feasibility study for
the 2-bin $p_T$-ratio in the context of TMD parameters determination.

\begin{figure*}[h]
    \centering
    \begin{subfigure}
        \centering
        \includegraphics[width=0.32\linewidth]{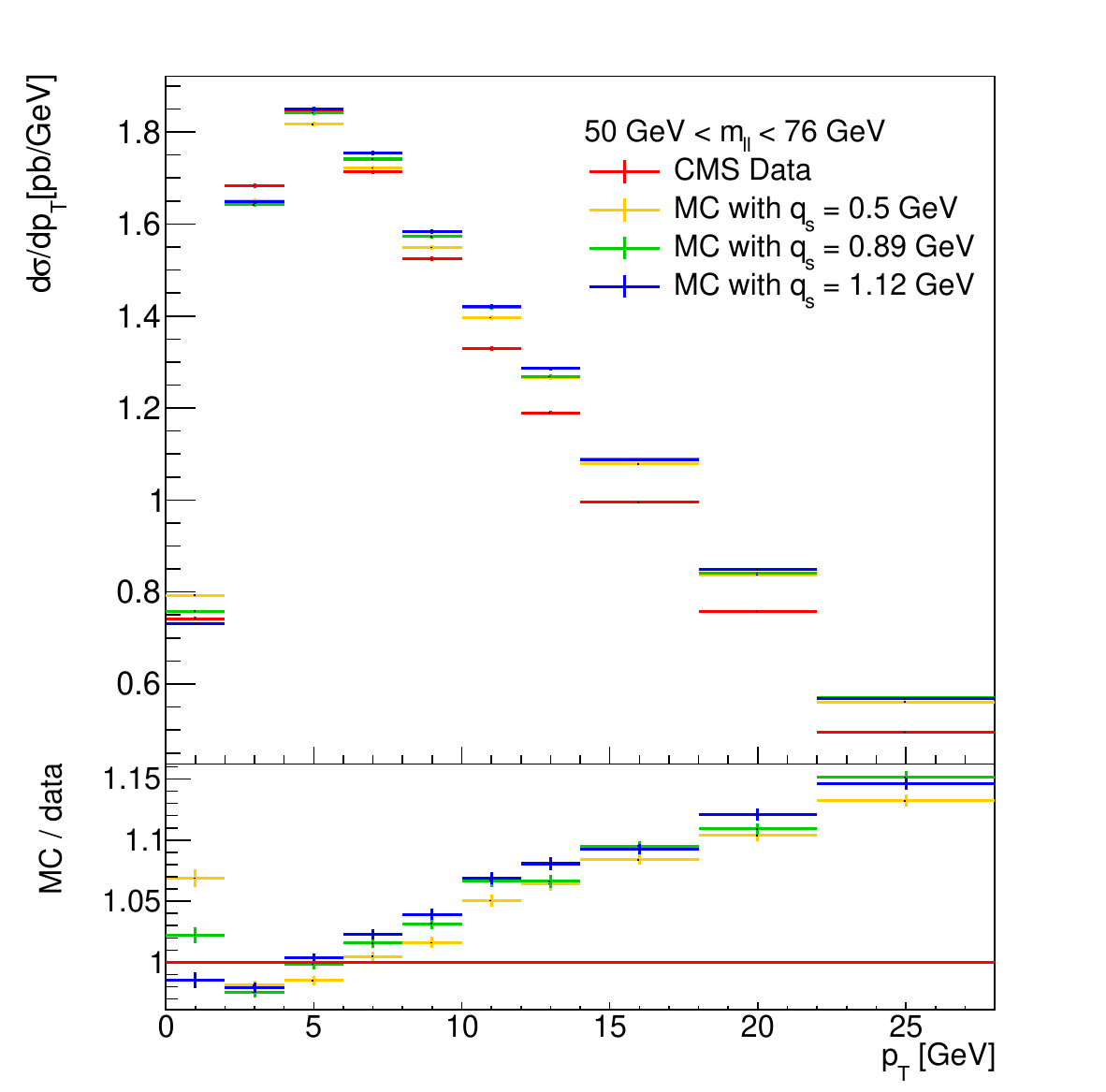}
    \end{subfigure}
    \begin{subfigure}
        \centering
        \includegraphics[width=0.32\linewidth]{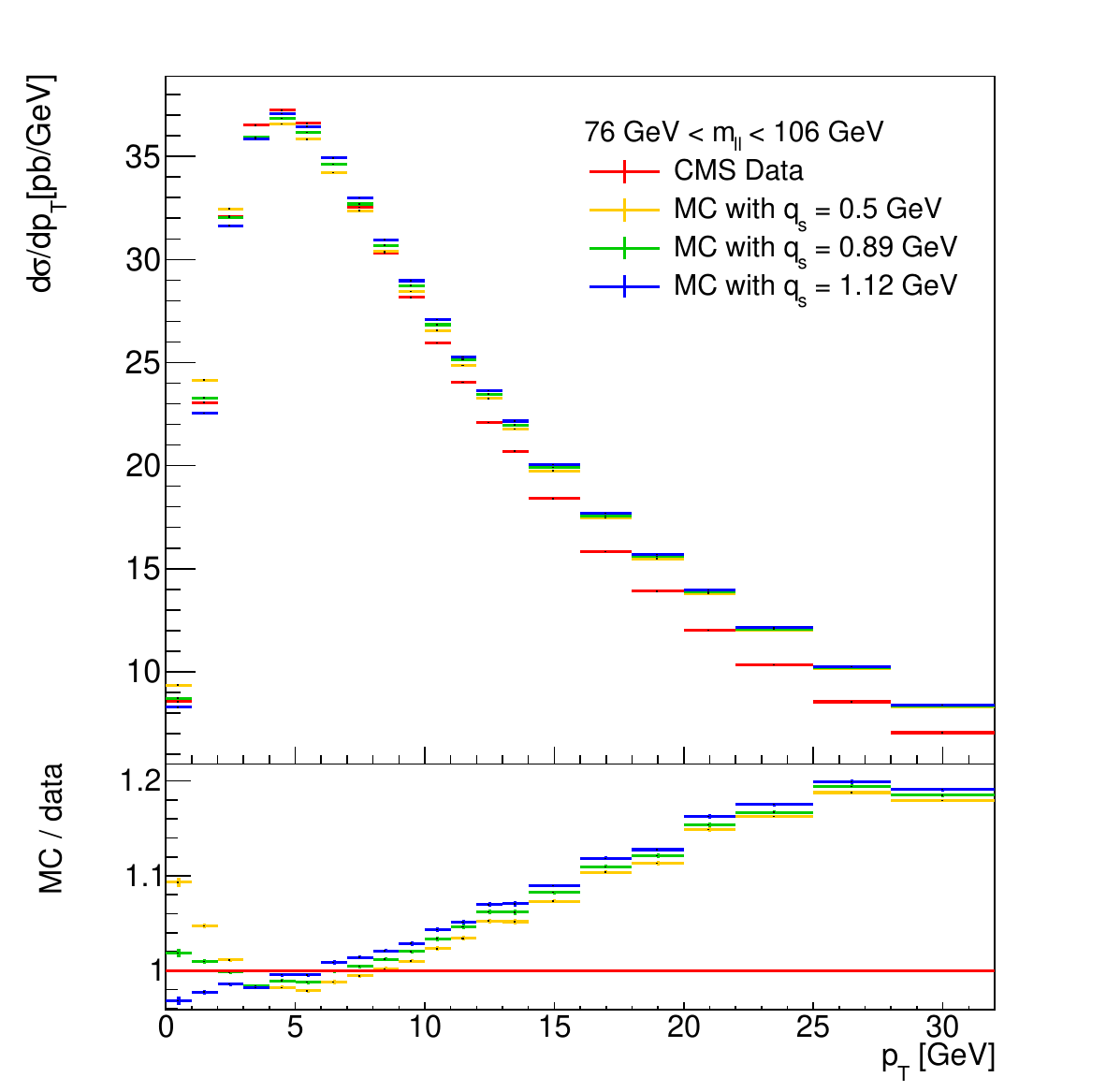}
    \end{subfigure}
    \begin{subfigure}
        \centering
        \includegraphics[width=0.32\linewidth]{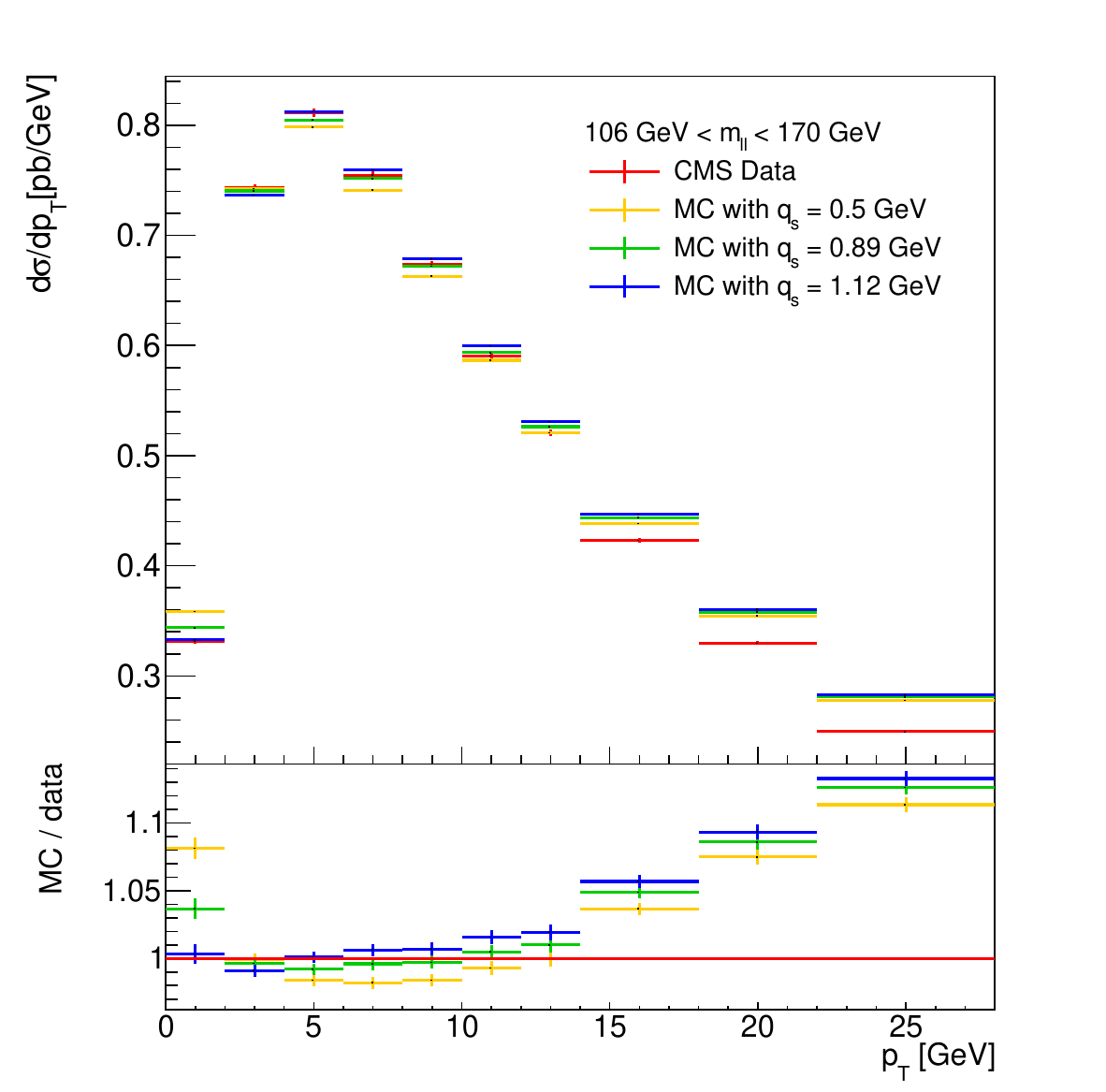}
    \end{subfigure}
    \begin{subfigure}
        \centering
        \includegraphics[width=0.32\linewidth]{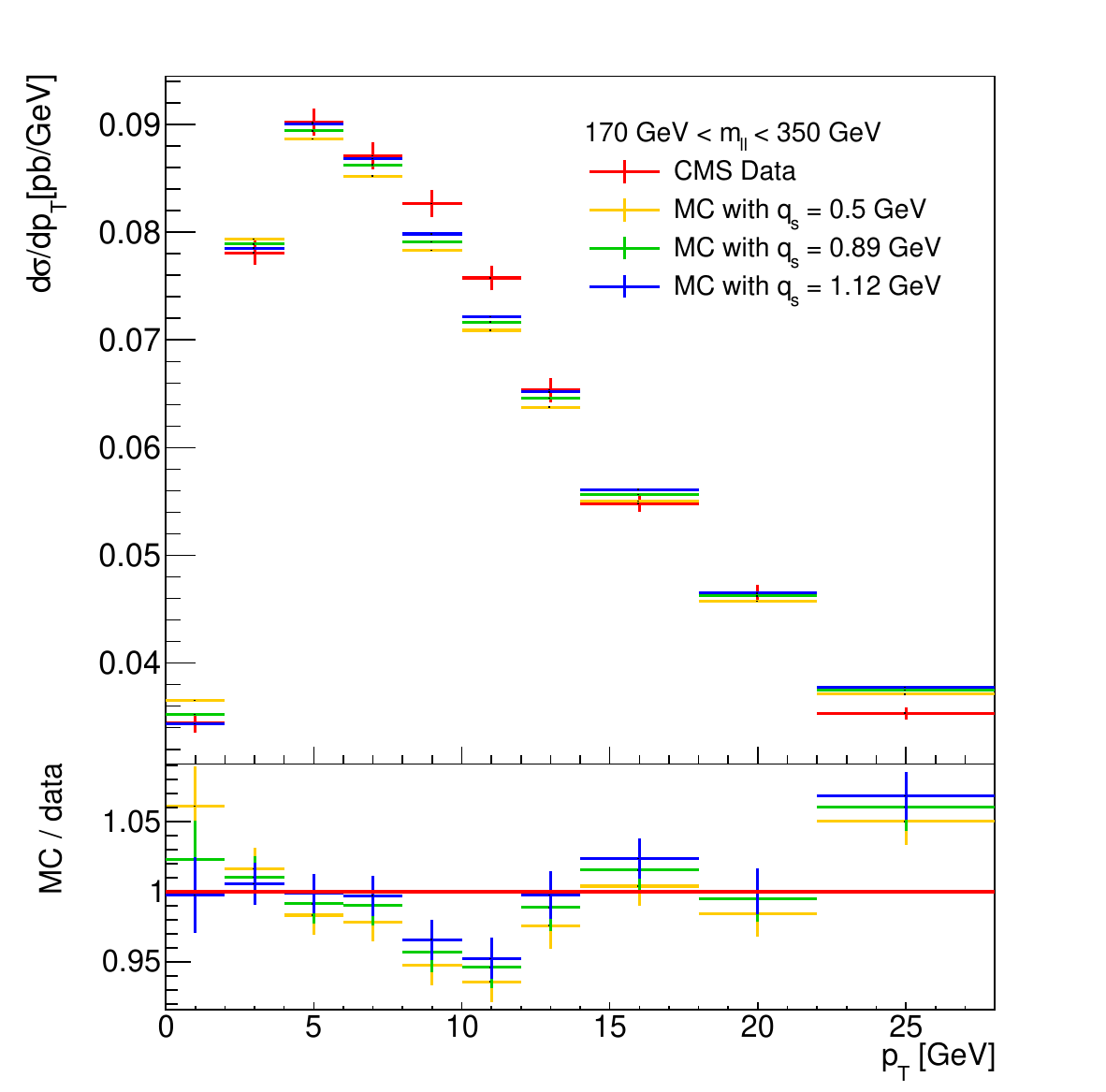}
    \end{subfigure}
    \begin{subfigure}
        \centering
        \includegraphics[width=0.32\linewidth]{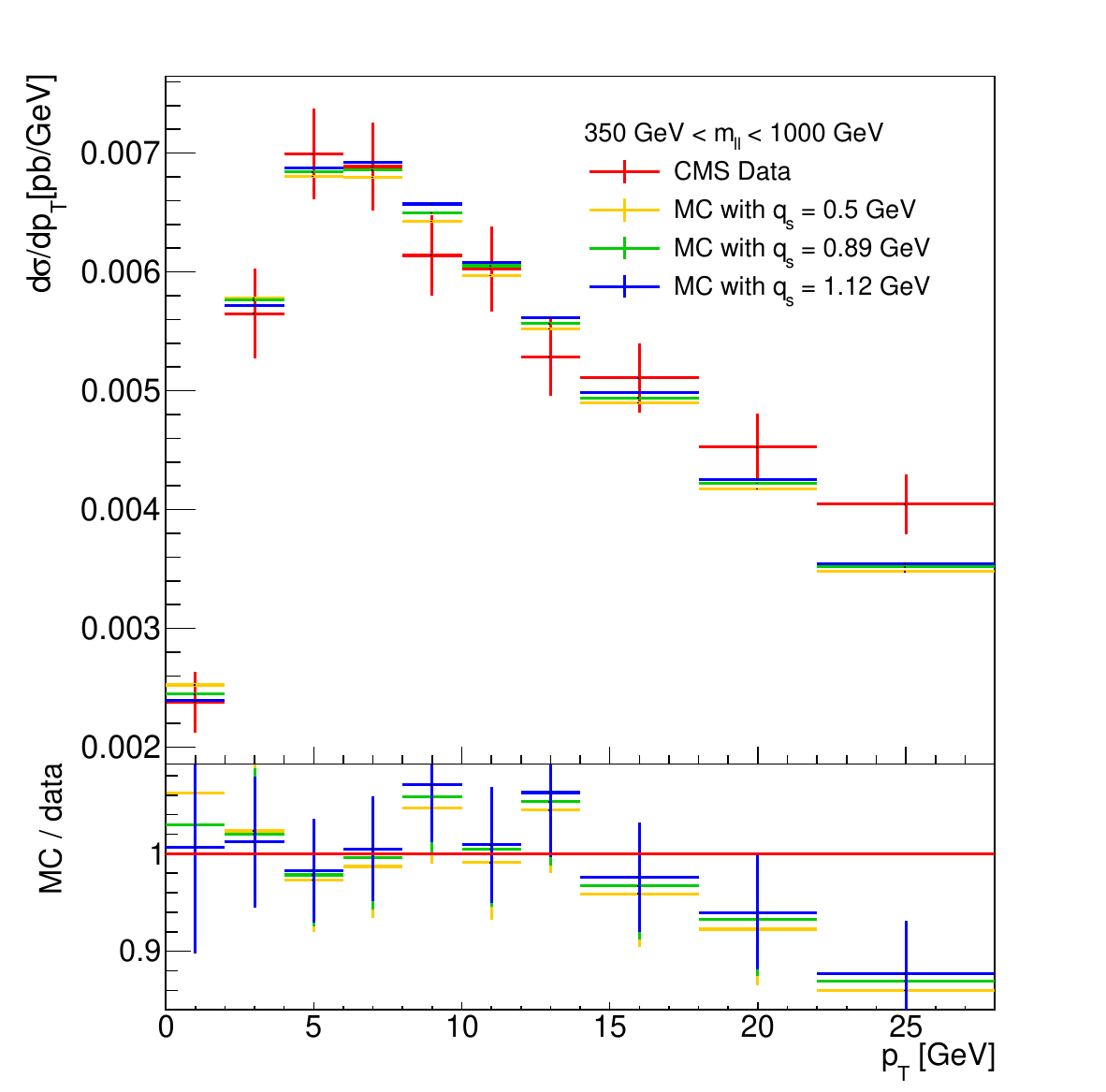}
    \end{subfigure}
    \caption{Comparison of $p_T$ distributions from the PB TMD samples with
    CMS data~\cite{CMS:2022ubq} in five $m_{ll}$ ranges. PB-NLO TMD Set2 with $q_s=0.5$ GeV, 0.89 GeV and 1.12 GeV are shown in the figure, as well as the measured data. The error bars represent the square root of the diagonal elements of the covariance matrix of model and data.}
    \label{fig:compare}
\end{figure*}

The CMS collaboration has measured the $p_T(ll)$ distribution in a wide mass range from 50 GeV to 1 TeV, with a bin width of 1 GeV in the DY invariant mass bin around the $Z$-boson  mass~\cite{CMS:2022ubq}. This measurement has been used  for the determination of the $q_s$  dependence on DY mass in Ref.~\cite{Bubanja:2023nrd}. We here perform an extraction of  $q_s$ from this measurement  using the  $p_T$-ratio, and compare the  results with those in Ref.~\cite{Bubanja:2023nrd}. Besides testing the feasibility of
the $p_T$-ratio proposal with real data, we also aim to check that the  $p_T$-ratio is not biased by the high-$p_T$ region, where contributions from higher jet
multiplicities are important~\cite{BermudezMartinez:2021lxz,BermudezMartinez:2022bpj}.

The selections of the leptons are
\begin{itemize}
    \item $|\eta|<2.4$ for both lepton,
    \item $p_T>25$ GeV for leading lepton,
    \item $p_T>20$ GeV for subleading lepton.
\end{itemize}
We use  dressed-level leptons, defined with $\Delta R<$ 0.1. The scale uncertainties are treated as fully correlated, and  computed by varying the renormalization and factorization scales by a factor of 2 separately,
excluding the two extreme cases. Since the model-induced statistical uncertainty is negligible comparing to the experimental uncertainty, the final covariance matrix $C_{ij}$ comprises total covariance matrix
from experiment
and scale uncertainty:
\begin{equation}
    C_{ij}=C_{ij}^{meas.}+\sigma_i^{\mu_{R/F}}\sigma_j^{\mu_{R/F}}
\end{equation}

The description of the CMS data by the theoretical predictions is shown in
Fig.~\ref{fig:compare}. In our study, the $p_T(ll)$ range is  the same as in
Ref.~\cite{Bubanja:2023nrd} -- 6 GeV for the first $m_{ll}$ bin, 7 GeV for the second bin, and 8 GeV for the rest. This limitation prevents the ratio from being biased by the difference between prediction and data in  the higher $p_T$ region. The choice of $p_s$ is largely restricted by the measurement, which affects the final sensitivity, as is suggested in the sensitivity study in Sect.~\ref{sec:ss}. We stress that, if
the $p_T$-ratio is used for  TMD parameter extractions, the choice of the
$p_s$ value will need careful investigation. For our test in this paper,
$p_s$  is set to 2 GeV for all $m_{DY}$ bins.

\begin{figure}[h]
    \centering
    \includegraphics[width=\linewidth]{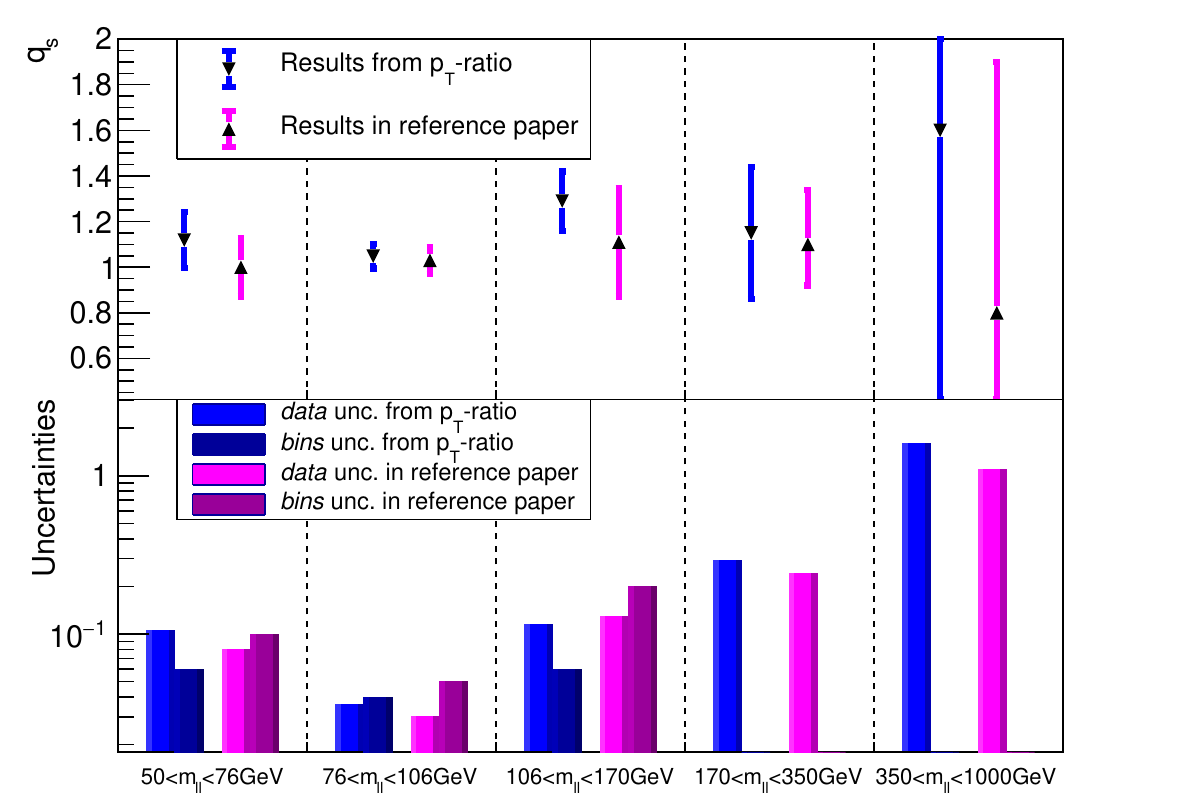}
    \caption{Determination of intrinsic $k_T$ parameter  $q_s$. The upper panel contains the $q_s$ values extracted independently in different $m_{DY}$ bins, compared with the results in Ref.~\cite{Bubanja:2023nrd}. The error bars represent the combinations of the corresponding uncertainties in the lower pad. The lower panel contains a comparison of  each source of uncertainties. In the last two bins we only consider \textit{data} uncertainties.}
    \label{fig:result}
\end{figure}

Final results are reported in Fig.~\ref{fig:result}. For comparison, the statistical uncertainties and scale uncertainties from data are considered, and compared to those in the
analysis~\cite{Bubanja:2023nrd}. The uncertainties labeled with \textit{data} at 68\% confidence level are obtained from the $q_s$ gap defined by $\chi^2=\chi^2_{min}+1$, and should be compared with the uncertainties labeled with \textit{data} in
Ref.~\cite{Bubanja:2023nrd}. Another source of uncertainties comes from the choice of the $p_T$ range. In Ref.~\cite{Bubanja:2023nrd}, this is estimated by varying the numbers of bins in the fit. In this study, we obtain it in the same way. For the last two $m_{DY}$ bins, it is not estimated for the lack of statistics. The result in every $m_{DY}$ region is consistent with the previous result~\cite{Bubanja:2023nrd}.

\section{Conclusion}\label{sec:conclusion}

In this paper we propose the $p_T(ll)$-ratio, which is defined as the ratio of low and high transverse momentum region of DY lepton pairs, as a sensitive observable and methodology for the extraction of a TMD parameter, the intrinsic $k_T$ Gaussian width $q_s$. We firstly review the basic idea of the parton branching method for TMD evolution, and the calculation of $p_T(ll)$ in the DY process. We then point out that the sensitivity of the $p_T(ll)$ shape to the intrinsic $k_T$ distribution actually lies in the shift of the spectrum from low $p_T$ to high $p_T$. This observation leads to the proposal of the $p_T(ll)$-ratio. We numerically test the sensitivity of the $p_T$-ratio to $q_s$ by comparing its statistical uncertainty with the same procedure on fine-binning $p_T$ structure. The result shows that the $p_T$-ratio has comparable sensitivity to the fine-binned $p_T$ shape.

The precision in the most sensitive $p_T(ll)$ regions in real measurements at the LHC is largely restricted by experimental systematic uncertainties. These make the fine-binned $p_T$  structure difficult to access experimentally, and point to the need to explore approaches based on coarse-grained binning. The relative ratio of low and high $p_T(ll)$ regions avoids the fine binning. In this paper, we illustrate its effectiveness by extracting $q_s$ from the $p_T$-ratio in a wide invariant-mass range of DY lepton pairs measured by the CMS collaboration recently. Since the CMS $p_T$ is binned into 2 GeV (or 1 GeV), we can only conduct this test by rebinning the distributions and covariance matrices, which means we cannot arbitrarily choose the separation momentum $p_s$. Nevertheless, the result is consistent with previous results obtained from the low-$p_T(ll)$ distribution of DY lepton pairs as a function of invariant mass, and has the same level of precision.

From this perspective, this methodology can be tested in future studies by using more complex TMD parameterization forms, or investigating extreme conditions of very small or very large $q_s$ values, to make sure the extraction is unbiased. It could be used in
forthcoming experiments for TMD performance studies, since it avoids the need to control large systematics in the low $p_T(ll)$ region. Unlike the extraction from a $p_T$ distribution that has been already binned, the $p_T$-ratio requires a dedicated study of the separation momentum $p_s$, affecting the sensitivity to TMD parameters, for example along the lines of the pseudo-data test carried out in this work.

\section*{Acknowledgement}

We are grateful to Hannes Jung for
discussion and for assistance with the use of PB TMD templates.
We  thank Laurent Favart, Jibo He, Hengne Li, Louis Moureaux and Hang Yin for
useful conversations.
This work was supported by the National Natural Science Foundation of China under Grants No.11721505, No.12061141005, and No.12105275, and supported by the ``USTC Research Funds of the Double First-Class Initiative".

\printbibliography

@article{RN66,
   author = {Sirunyan, A. M. and Tumasyan, A. and Adam, W. and Ambrogi, F. and Bergauer, T. and Brandstetter, J. and Dragicevic, M. and Erö, J. and Escalante Del Valle, A. and Flechl, M. and Frühwirth, R. and Jeitler, M. and Krammer, N. and Krätschmer, I. and Liko, D. and Madlener, T. and Mikulec, I. and Rad, N. and Schieck, J. and Schöfbeck, R. and Spanring, M. and Spitzbart, D. and Waltenberger, W. and Wulz, C. E. and Zarucki, M. and Drugakov, V. and Mossolov, V. and Suarez Gonzalez, J. and Darwish, M. R. and De Wolf, E. A. and Di Croce, D. and Janssen, X. and Lelek, A. and Pieters, M. and Rejeb Sfar, H. and Van Haevermaet, H. and Van Mechelen, P. and Van Putte, S. and Van Remortel, N. and Blekman, F. and Bols, E. S. and Chhibra, S. S. and D'Hondt, J. and De Clercq, J. and Lontkovskyi, D. and Lowette, S. and Marchesini, I. and Moortgat, S. and Python, Q. and Skovpen, K. and Tavernier, S. and Van Doninck, W. and Van Mulders, P. and Beghin, D. and Bilin, B. and Brun, H. and Clerbaux, B. and De Lentdecker, G. and Delannoy, H. and Dorney, B. and Favart, L. and Grebenyuk, A. and Kalsi, A. K. and Popov, A. and Postiau, N. and Starling, E. and Thomas, L. and Vander Velde, C. and Vanlaer, P. and Vannerom, D. and Cornelis, T. and Dobur, D. and Khvastunov, I. and Niedziela, M. and Roskas, C. and Trocino, D. and Tytgat, M. and Verbeke, W. and Vermassen, B. and Vit, M. and Bondu, O. and Bruno, G. and Caputo, C. and David, P. and Delaere, C. and Delcourt, M. and Giammanco, A. and Lemaitre, V. and Prisciandaro, J. and Saggio, A. and Vidal Marono, M. and Vischia, P. and Zobec, J. and Alves, F. L. and Alves, G. A. and Correia Silva, G. and Hensel, C. and Moraes, A. and Rebello Teles, P. and Belchior Batista Das Chagas, E. and others },
   title = {Measurements of differential Z boson production cross sections in proton-proton collisions at $\sqrt{s}$ = 13 TeV},
   journal = {Journal of High Energy Physics},
   volume = {2019},
   number = {12},
   ISSN = {1029-8479},
   DOI = {10.1007/jhep12(2019)061},
   year = {2019},
   type = {Journal Article}
}

@article{RN60,
   author = {Aad, G. and Abbott, B. and Abbott, D. C. and Abud, A. Abed and Abeling, K. and Abhayasinghe, D. K. and Abidi, S. H. and AbouZeid, O. S. and Abraham, N. L. and Abramowicz, H. and Abreu, H. and Abulaiti, Y. and Acharya, B. S. and Achkar, B. and Adachi, S. and Adam, L. and Bourdarios, C. Adam and Adamczyk, L. and Adamek, L. and Adelman, J. and Adersberger, M. and Adiguzel, A. and Adorni, S. and Adye, T. and Affolder, A. A. and Afik, Y. and Agapopoulou, C. and Agaras, M. N. and Aggarwal, A. and Agheorghiesei, C. and Aguilar-Saavedra, J. A. and Ahmadov, F. and Ahmed, W. S. and Ai, X. and Aielli, G. and Akatsuka, S. and Åkesson, T. P. A. and Akilli, E. and Akimov, A. V. and Khoury, K. Al and Alberghi, G. L. and Albert, J. and Verzini, M. J. Alconada and Alderweireldt, S. and Aleksa, M. and Aleksandrov, I. N. and Alexa, C. and Alexopoulos, T. and Alfonsi, A. and Alfonsi, F. and Alhroob, M. and Ali, B. and Aliev, M. and Alimonti, G. and Alkire, S. P. and Allaire, C. and Allbrooke, B. M. M. and Allen, B. W. and Allport, P. P. and Aloisio, A. and Alonso, F. and Alpigiani, C. and Alshehri, A. A. and Camelia, E. Alunno and Estevez, M. Alvarez and Alviggi, M. G. and Coutinho, Y. Amaral and Ambler, A. and Ambroz, L. and Amelung, C. and Amidei, D. and Santos, S. P. Amor Dos and Amoroso, S. and Amrouche, C. S. and An, F. and Anastopoulos, C. and Andari, N. and Andeen, T. and Anders, C. F. and Anders, J. K. and Andreazza, A. and Andrei, V. and Anelli, C. R. and Angelidakis, S. and Angerami, A. and Anisenkov, A. V. and Annovi, A. and Antel, C. and Anthony, M. T. and Antipov, E. and Antonelli, M. and Antrim, D. J. A. and Anulli, F. and Aoki, M. and Pozo, J. A. Aparisi and Bella, L. Aperio and Araque, J. P. and Ferraz, V. Araujo and Pereira, R. Araujo and Arcangeletti, C. and others },
   title = {Measurement of the transverse momentum distribution of Drell-Yan lepton pairs in proton-proton collisions at $\sqrt{s}=13$TeV with the ATLAS detector},
   journal = {The European Physical Journal C},
   volume = {80},
   number = {7},
   ISSN = {1434-6044
1434-6052},
   DOI = {10.1140/epjc/s10052-020-8001-z},
   year = {2020},
   type = {Journal Article}
}

@article{pythia,
   doi = {10.1088/1126-6708/2006/05/026},
   url = {https://dx.doi.org/10.1088/1126-6708/2006/05/026},
   year = {2006},
   month = {5},
   publisher = {},
   volume = {2006},
   number = {05},
   pages = {026},
   author = {Torbjorn Sjostrand and  Stephen Mrenna and  Peter Skands}, 
   title = {PYTHIA 6.4 physics and manual},
   journal = {Journal of High Energy Physics}
}

@article{GennaroCorcella_2001,
doi = {10.1088/1126-6708/2001/01/010},
url = {https://dx.doi.org/10.1088/1126-6708/2001/01/010},
year = {2001},
month = {2},
publisher = {},
volume = {2001},
number = {01},
pages = {010},
author = {Gennaro Corcella and  Ian G. Knowles and  Giuseppe Marchesini and  Stefano Moretti and  Kosuke Odagiri and  Peter Richardson and  Michael H. Seymour and  Bryan R. Webber},
title = {HERWIG 6: an event generator for hadron emission reactions with interfering gluons (including supersymmetric processes)},
journal = {Journal of High Energy Physics}
}

@article{Corcella:2002jc,
    author = "Corcella, G. and Knowles, I. G. and Marchesini, G. and Moretti, S. and Odagiri, K. and Richardson, P. and Seymour, M. H. and Webber, B. R.",
    title = "{HERWIG 6.5 release note}",
    eprint = "hep-ph/0210213",
    archivePrefix = "arXiv",
    reportNumber = "CAVENDISH-HEP-02-17, DAMTP-2002-124, KEK-TH-850, MPI-PHT-2002-55, CERN-TH-2002-270, IPPP-02-58, MC-TH-2002-7",
    month = "10",
    year = "2002"
}

@inproceedings{Monfared:2024vgc,
    author = "Monfared, S. Taheri",
    title = "{Recent progress in transverse momentum dependent (TMD) Parton Densities and corresponding parton showers}",
    booktitle = "{31st International Workshop on Deep-Inelastic Scattering and Related Subjects}",
    eprint = "2410.05853",
    archivePrefix = "arXiv",
    primaryClass = "hep-ph",
    reportNumber = "DESY-24-148",
    month = "10",
    year = "2024"
}

@inproceedings{Lelek:2024kax,
    author = "Lelek, Aleksandra",
    title = "{NNLL Transverse Momentum Dependent evolution in the Parton Branching method}",
    booktitle = "{42nd International Symposium on Physics In Collision}",
    eprint = "2412.09108",
    archivePrefix = "arXiv",
    primaryClass = "hep-ph",
    month = "12",
    year = "2024"
}

@article{Gribov:1972ri,
    author = "Gribov, V. N. and Lipatov, L. N.",
    title = "{Deep inelastic e p scattering in perturbation theory}",
    reportNumber = "IPTI-381-71",
    journal = "Sov. J. Nucl. Phys.",
    volume = "15",
    pages = "438--450",
    year = "1972"
}

@article{BermudezMartinez:2020tys,
    author = "Berm{\'u}dez Mart{\'i}nez, A. and others",
    title = "{The transverse momentum spectrum of low mass Drell\textendash{}Yan production at next-to-leading order in the parton branching method}",
    eprint = "2001.06488",
    archivePrefix = "arXiv",
    primaryClass = "hep-ph",
    reportNumber = "DESY 20-006, DESY-20-006",
    doi = "10.1140/epjc/s10052-020-8136-y",
    journal = "Eur. Phys. J. C",
    volume = "80",
    number = "7",
    pages = "598",
    year = "2020"
}

@Article{Alwall:2014hca,
  author        = {Alwall, J. and Frederix, R. and Frixione, S. and Hirschi, V. and Maltoni, F. and Mattelaer, O. and Shao, H. -S. and Stelzer, T. and Torrielli, P. and Zaro, M.},
  title         = {{The automated computation of tree-level and next-to-leading order differential cross sections, and their matching to parton shower simulations}},
  journal       = {JHEP},
  year          = {2014},
  volume        = {07},
  pages         = {079},
  archiveprefix = {arXiv},
  doi           = {10.1007/JHEP07(2014)079},
  eprint        = {1405.0301},
  primaryclass  = {hep-ph},
  reportnumber  = {CERN-PH-TH-2014-064, CP3-14-18, LPN14-066, MCNET-14-09, ZU-TH-14-14},
}

@Article{Amati:1980ch,
  Title                    = {{A Treatment of Hard Processes Sensitive to the Infrared Structure of QCD}},
  Author                   = {Amati, D. and Bassetto, A. and Ciafaloni, M. and Marchesini, G. and Veneziano, G.},
  Journal                  = {Nucl. Phys.},
  Year                     = {1980},
  Pages                    = {429-455},
  Volume                   = {B173},

  Doi                      = {10.1016/0550-3213(80)90012-7},
  Reportnumber             = {CERN-TH-2831},
  Slaccitation             = {%%CITATION = NUPHA,B173,429;%%}
}

@article{Bassetto:1983mvz,
    author = "Bassetto, A. and Ciafaloni, M. and Marchesini, G.",
    title = "{Jet Structure and Infrared Sensitive Quantities in Perturbative QCD}",
    doi = "10.1016/0370-1573(83)90083-2",
    journal = "Phys. Rept.",
    volume = "100",
    pages = "201--272",
    year = "1983"
}

@inproceedings{Barzani:2022msy,
    author = "Sadeghi Barzani, S.",     
    title = "{PB TMD fits at NLO with dynamical resolution scale}",
    booktitle = "{29th International Workshop on Deep-Inelastic Scattering and Related Subjects}",
    eprint = "2207.13519",
    archivePrefix = "arXiv",
    primaryClass = "hep-ph",
    month = "7",
    year = "2022"
}

@article{CMS:2022ubq,
    author = "Tumasyan, Armen and others",
    collaboration = "CMS",
    title = "{Measurement of the mass dependence of the transverse momentum of lepton pairs in Drell-Yan production in proton-proton collisions at $\sqrt{s}$ = 13 TeV}",
    eprint = "2205.04897",
    archivePrefix = "arXiv",
    primaryClass = "hep-ex",
    reportNumber = "CMS-SMP-20-003, CERN-EP-2022-053",
    doi = "10.1140/epjc/s10052-023-11631-7",
    journal = "Eur. Phys. J. C",
    volume = "83",
    number = "7",
    pages = "628",
    year = "2023"
}

@article{LHCb:2021huf,
    author = "Aaij, R. and others",
    collaboration = "LHCb",
    title = "{Precision measurement of forward $Z$ boson production in proton-proton collisions at $\sqrt{s} = 13$ TeV}",
    eprint = "2112.07458",
    archivePrefix = "arXiv",
    primaryClass = "hep-ex",
    reportNumber = "LHCb-PAPER-2021-037, CERN-EP-2021-246",
    doi = "10.1007/JHEP07(2022)026",
    journal = "JHEP",
    volume = "07",
    pages = "026",
    year = "2022"
}

@article{ATLAS:2015iiu,
    author = "Aad, Georges and others",
    collaboration = "ATLAS",
    title = "{Measurement of the transverse momentum and $\phi ^*_{\eta }$ distributions of Drell\textendash{}Yan lepton pairs in proton\textendash{}proton collisions at $\sqrt{s}=8$  TeV with the ATLAS detector}",
    eprint = "1512.02192",
    archivePrefix = "arXiv",
    primaryClass = "hep-ex",
    reportNumber = "CERN-PH-EP-2015-275",
    doi = "10.1140/epjc/s10052-016-4070-4",
    journal = "Eur. Phys. J. C",
    volume = "76",
    number = "5",
    pages = "291",
    year = "2016"
}

@article{D0:1999jba,
    author = "Abbott, B. and others",
    collaboration = "D0",
    title = "{Measurement of the inclusive differential cross section for $Z$ bosons as a function of transverse momentum in $\bar{p}p$ collisions at $\sqrt{s} = 1.8$ TeV}",
    eprint = "hep-ex/9907009",
    archivePrefix = "arXiv",
    reportNumber = "FERMILAB-PUB-99-197-E",
    doi = "10.1103/PhysRevD.61.032004",
    journal = "Phys. Rev. D",
    volume = "61",
    pages = "032004",
    year = "2000"
}

@article{CDF:1999bpw,
    author = "Affolder, T. and others",
    collaboration = "CDF",
    title = "{The transverse momentum and total cross section of $e^+e^-$ pairs in the $Z$ boson region from $p\bar{p}$ collisions at $\sqrt{s} = 1.8$ TeV}",
    eprint = "hep-ex/0001021",
    archivePrefix = "arXiv",
    reportNumber = "FERMILAB-PUB-99-220-E",
    doi = "10.1103/PhysRevLett.84.845",
    journal = "Phys. Rev. Lett.",
    volume = "84",
    pages = "845--850",
    year = "2000"
}

@article{CDF:2012brb,
    author = "Aaltonen, T. and others",
    collaboration = "CDF",
    title = "{Transverse momentum cross section of $e^+e^-$ pairs in the $Z$-boson region from $p\bar{p}$ collisions at $\sqrt{s}=1.96$ TeV}",
    eprint = "1207.7138",
    archivePrefix = "arXiv",
    primaryClass = "hep-ex",
    reportNumber = "FERMILAB-PUB-12-421-E",
    doi = "10.1103/PhysRevD.86.052010",
    journal = "Phys. Rev. D",
    volume = "86",
    pages = "052010",
    year = "2012"
}

@article{PHENIX:2018dwt,
    author = "Aidala, C. and others",
    collaboration = "PHENIX",
    title = "{Measurements of $\mu\mu$ pairs from open heavy flavor and Drell-Yan in $p+p$ collisions at $\sqrt{s}=200$ GeV}",
    eprint = "1805.02448",
    archivePrefix = "arXiv",
    primaryClass = "hep-ex",
    doi = "10.1103/PhysRevD.99.072003",
    journal = "Phys. Rev. D",
    volume = "99",
    number = "7",
    pages = "072003",
    year = "2019"
}

@article{CMS:2021ynu,
    author = "Sirunyan, Albert M and others",
    collaboration = "CMS",
    title = "{Study of Drell-Yan dimuon production in proton-lead collisions at $\sqrt{s_\mathrm{NN}} =$ 8.16 TeV}",
    eprint = "2102.13648",
    archivePrefix = "arXiv",
    primaryClass = "hep-ex",
    reportNumber = "CMS-HIN-18-003, CERN-EP-2021-028",
    doi = "10.1007/JHEP05(2021)182",
    journal = "JHEP",
    volume = "05",
    pages = "182",
    year = "2021"
}

@article{Moreno:1990sf,
    author = "Moreno, G. and others",
    title = "{Dimuon Production in Proton - Copper Collisions at $\sqrt{s}$ = 38.8-GeV}",
    reportNumber = "FERMILAB-PUB-90-223-E",
    doi = "10.1103/PhysRevD.43.2815",
    journal = "Phys. Rev. D",
    volume = "43",
    pages = "2815--2836",
    year = "1991"
}

@Article{Catani:1990rr,
  author       = {Catani, S. and Webber, B. R. and Marchesini, G.},
  title        = {{QCD coherent branching and semiinclusive processes at large x}},
  journal      = {Nucl. Phys.},
  year         = {1991},
  volume       = {B349},
  pages        = {635-654},
  doi          = {10.1016/0550-3213(91)90390-J},
  reportnumber = {CAVENDISH-HEP-90-11, UPRF-90-280},
  slaccitation = {%%CITATION = NUPHA,B349,635;%%},
}

@Article{Alekhin:2014irh,
  author        = {Alekhin, S. and others},
  title         = {{HERAFitter}},
  journal       = {Eur. Phys. J. C},
  year          = {2015},
  volume        = {75},
  number        = {7},
  pages         = {304},
  archiveprefix = {arXiv},
  doi           = {10.1140/epjc/s10052-015-3480-z},
  eprint        = {1410.4412},
  primaryclass  = {hep-ph},
  reportnumber  = {DESY-14-188, DESY-REPORT-14-188, FERMILAB-PUB-14-603-CMS},
}

@article{CASCADE:2010clj,
    author = "Jung, H. and others",
    collaboration = "CASCADE",
    title = "{The CCFM Monte Carlo generator CASCADE version 2.2.03}",
    eprint = "1008.0152",
    archivePrefix = "arXiv",
    primaryClass = "hep-ph",
    reportNumber = "DESY-10-107",
    doi = "10.1140/epjc/s10052-010-1507-z",
    journal = "Eur. Phys. J. C",
    volume = "70",
    pages = "1237--1249",
    year = "2010"
}

@article{Martinez:2024twn,
    author = "Bermudez Martinez, A.  and others",
    title = "{The Parton Branching Sudakov and its relation to CSS}",
    doi = "10.22323/1.449.0270",
    journal = "PoS",
    volume = "EPS-HEP2023",
    pages = "270",
    year = "2024"
}

@article{CMS:2024eprint,
    author = "Hayrapetyan, A. and others",
    collaboration = "CMS",    
    title = "{Energy scaling behavior of intrinsic transverse momentum parameters in Drell-Yan simulation}",
    eprint = "2409.17770",
    archivePrefix = "arXiv",
    primaryClass = "hep-ph",
    month = "9",
    year = "2024"
}

@article{Bubanja:2024puv,
    author = "Bubanja, I. and Jung, H. and Lelek, A. and Raicevic, N. and Taheri Monfared, S.",
    title = "{Center-of-mass energy dependence of intrinsic-$k_T$ distributions obtained from Drell-Yan production}",
    eprint = "2404.04088",
    archivePrefix = "arXiv",
    primaryClass = "hep-ph",
    reportNumber = "DESY-24-049",
    month = "4",
    year = "2024"
}

@inproceedings{Jung:2024eam,
    author = "Jung, H.",
    title = "{The non-perturbative Sudakov Form Factor and the role of soft gluons}",
    booktitle = "{30th Cracow Epiphany Conference on on Precision Physics at High Energy Colliders}: {dedicated to the memory of Staszek Jadach}",
    eprint = "2404.06905",
    archivePrefix = "arXiv",
    primaryClass = "hep-ph",
    month = "4",
    year = "2024"
}

@article{Abramowicz:2015mha,
    author = "Abramowicz, H. and others",
    collaboration = "H1, ZEUS",
    title = "{Combination of measurements of inclusive deep inelastic ${e^{\pm }p}$ scattering cross sections and QCD analysis of HERA data}",
    eprint = "1506.06042",
    archivePrefix = "arXiv",
    primaryClass = "hep-ex",
    reportNumber = "DESY-15-039",
    doi = "10.1140/epjc/s10052-015-3710-4",
    journal = "Eur. Phys. J. C",
    volume = "75",
    number = "12",
    pages = "580",
    year = "2015"
}

@article{Altarelli:1977zs,
    author = "Altarelli, Guido and Parisi, G.",
    title = "{Asymptotic Freedom in Parton Language}",
    reportNumber = "LPTENS-77-6",
    doi = "10.1016/0550-3213(77)90384-4",
    journal = "Nucl. Phys. B",
    volume = "126",
    pages = "298--318",
    year = "1977"
}

@article{Dokshitzer:1977sg,
    author = "Dokshitzer, Yuri L.",
    title = "{Calculation of the Structure Functions for Deep Inelastic Scattering and e+ e- Annihilation by Perturbation Theory in Quantum Chromodynamics.}",
    journal = "Sov. Phys. JETP",
    volume = "46",
    pages = "641--653",
    year = "1977"
}

@article{Angeles-Martinez:2015sea,
    author = "Angeles-Martinez, R. and others",
    title = "{Transverse Momentum Dependent (TMD) parton distribution functions: status and prospects}",
    eprint = "1507.05267",
    archivePrefix = "arXiv",
    primaryClass = "hep-ph",
    reportNumber = "DESY-15-111, NIKHEF-2015-023, RAL-P-2015-006, JLAB-THY-15-2020",
    doi = "10.5506/APhysPolB.46.2501",
    journal = "Acta Phys. Polon. B",
    volume = "46",
    number = "12",
    pages = "2501--2534",
    year = "2015"
}

@article{Bacchetta:2024qre,
    author = "Bacchetta, Alessandro and Bertone, Valerio and Bissolotti, Chiara and Bozzi, Giuseppe and Cerutti, Matteo and Delcarro, Filippo and Radici, Marco and Rossi, Lorenzo and Signori, Andrea",
    collaboration = "MAP",
    title = "{Flavor dependence of unpolarized quark transverse momentum distributions from a global fit}",
    eprint = "2405.13833",
    archivePrefix = "arXiv",
    primaryClass = "hep-ph",
    reportNumber = "JLAB-THY-24-4066",
    doi = "10.1007/JHEP08(2024)232",
    journal = "JHEP",
    volume = "08",
    pages = "232",
    year = "2024"
}

@article{Bacchetta:2022awv,
    author = "Bacchetta, Alessandro and Bertone, Valerio and Bissolotti, Chiara and Bozzi, Giuseppe and Cerutti, Matteo and Piacenza, Fulvio and Radici, Marco and Signori, Andrea",
    collaboration = "MAP (Multi-dimensional Analyses of Partonic distributions)",
    title = "{Unpolarized transverse momentum distributions from a global fit of Drell-Yan and semi-inclusive deep-inelastic scattering data}",
    eprint = "2206.07598",
    archivePrefix = "arXiv",
    primaryClass = "hep-ph",
    doi = "10.1007/JHEP10(2022)127",
    journal = "JHEP",
    volume = "10",
    pages = "127",
    year = "2022"
}

@article{Moos:2023yfa,
    author = "Moos, Valentin and Scimemi, Ignazio and Vladimirov, Alexey and Zurita, Pia",
    title = "{Extraction of unpolarized transverse momentum distributions from the fit of Drell-Yan data at N$^{4}$LL}",
    eprint = "2305.07473",
    archivePrefix = "arXiv",
    primaryClass = "hep-ph",
    reportNumber = "IPARCOS-UCM-035",
    doi = "10.1007/JHEP05(2024)036",
    journal = "JHEP",
    volume = "05",
    pages = "036",
    year = "2024"
}

@article{Bury:2022czx,
    author = "Bury, Marcin and Hautmann, Francesco and Leal-Gomez, Sergio and Scimemi, Ignazio and Vladimirov, Alexey and Zurita, Pia",
    title = "{PDF bias and flavor dependence in TMD distributions}",
    eprint = "2201.07114",
    archivePrefix = "arXiv",
    primaryClass = "hep-ph",
    reportNumber = "UWThPh 2021-29, CERN-TH-2022-126",
    doi = "10.1007/JHEP10(2022)118",
    journal = "JHEP",
    volume = "10",
    pages = "118",
    year = "2022"
}

@article{Camarda:2022qdg,
    author = "Camarda, Stefano and Ferrera, Giancarlo and Schott, Matthias",
    title = "{Determination of the strong-coupling constant from the Z-boson transverse-momentum distribution}",
    eprint = "2203.05394",
    archivePrefix = "arXiv",
    primaryClass = "hep-ph",
    doi = "10.1140/epjc/s10052-023-12373-2",
    journal = "Eur. Phys. J. C",
    volume = "84",
    number = "1",
    pages = "39",
    year = "2024"
}

@article{BermudezMartinez:2019anj,
    author = "Bermudez Martinez, A. and others",
    title = "{Production of Z-bosons in the parton branching method}",
    eprint = "1906.00919",
    archivePrefix = "arXiv",
    primaryClass = "hep-ph",
    reportNumber = "DESY-19-087, CERN-TH-2019-095, DESY 19-087",
    doi = "10.1103/PhysRevD.100.074027",
    journal = "Phys. Rev. D",
    volume = "100",
    number = "7",
    pages = "074027",
    year = "2019"
}

@article{BermudezMartinez:2021lxz,
    author = "Bermudez Martinez, A. and Hautmann, F. and Mangano, M. L.",
    title = "{TMD evolution and multi-jet merging}",
    eprint = "2107.01224",
    archivePrefix = "arXiv",
    primaryClass = "hep-ph",
    doi = "10.1016/j.physletb.2021.136700",
    journal = "Phys. Lett. B",
    volume = "822",
    pages = "136700",
    year = "2021"
}

@article{BermudezMartinez:2022bpj,
    author = "Bermudez Martinez, A. and Hautmann, F. and Mangano, M. L.",
    title = "{Multi-jet merging with TMD parton branching}",
    eprint = "2208.02276",
    archivePrefix = "arXiv",
    primaryClass = "hep-ph",
    reportNumber = "CERN-TH-2022-131, DESY-22-133",
    doi = "10.1007/JHEP09(2022)060",
    journal = "JHEP",
    volume = "09",
    pages = "060",
    year = "2022"
}

@article{Bubanja:2023nrd,
    author = "Bubanja, I. and others",
    title = "{The small $k_{\textrm{T}}$ region in Drell-Yan production at next-to-leading order with the parton branching method}",
    eprint = "2312.08655",
    archivePrefix = "arXiv",
    primaryClass = "hep-ph",
    reportNumber = "DESY-23-209",
    doi = "10.1140/epjc/s10052-024-12507-0",
    journal = "Eur. Phys. J. C",
    volume = "84",
    number = "2",
    pages = "154",
    year = "2024"
}

@article{Hautmann:2019biw,
    author = "Hautmann, F. and Keersmaekers, L. and Lelek, A. and Van Kampen, A. M.",
    title = "{Dynamical resolution scale in transverse momentum distributions at the LHC}",
    eprint = "1908.08524",
    archivePrefix = "arXiv",
    primaryClass = "hep-ph",
    reportNumber = "CERN-TH-2019-130",
    doi = "10.1016/j.nuclphysb.2019.114795",
    journal = "Nucl. Phys. B",
    volume = "949",
    pages = "114795",
    year = "2019"
}

@article{Collins:1984kg,
    author = "Collins, John C. and Soper, Davison E. and Sterman, George F.",
    title = "{Transverse Momentum Distribution in Drell-Yan Pair and W and Z Boson Production}",
    reportNumber = "CERN-TH-3923",
    doi = "10.1016/0550-3213(85)90479-1",
    journal = "Nucl. Phys. B",
    volume = "250",
    pages = "199--224",
    year = "1985"
}

@book{Collins:2011zzd,
    author = "Collins, John",
    title = "{Foundations of Perturbative QCD}",
    doi = "10.1017/9781009401845",
    isbn = "978-1-00-940184-5, 978-1-00-940183-8, 978-1-00-940182-1",
    publisher = "Cambridge University Press",
    series = "Cambridge Monographs on Particle Physics, Nuclear Physics and Cosmology",
    volume = "32",
    month = "7",
    year = "2023"
}

@article{Hautmann:2020cyp,
    author = "Hautmann, Francesco and Scimemi, Ignazio and Vladimirov, Alexey",
    title = "{Non-perturbative contributions to vector-boson transverse momentum spectra in hadronic collisions}",
    eprint = "2002.12810",
    archivePrefix = "arXiv",
    primaryClass = "hep-ph",
    doi = "10.1016/j.physletb.2020.135478",
    journal = "Phys. Lett. B",
    volume = "806",
    pages = "135478",
    year = "2020"
}

@article{Hautmann:2021ovt,
    author = "Hautmann, Francesco and Scimemi, Ignazio and Vladimirov, Alexey",
    title = "{Determination of the rapidity evolution kernel from Drell-Yan data at low transverse momenta}",
    eprint = "2109.12051",
    archivePrefix = "arXiv",
    primaryClass = "hep-ph",
    doi = "10.21468/SciPostPhysProc.8.123",
    journal = "SciPost Phys. Proc.",
    volume = "8",
    pages = "123",
    year = "2022"
}

@Article{Dooling:2012uw,
  Title                    = {{Longitudinal momentum shifts, showering, and nonperturbative corrections in matched next-to-leading-order shower event generators}},
  Author                   = {Dooling, S. and Gunnellini, P. and Hautmann, F. and Jung, H.},
  Journal                  = {Phys. Rev.},
  Year                     = {2013},
  Number                   = {9},
  Pages                    = {094009},
  Volume                   = {D87},

  Archiveprefix            = {arXiv},
  Doi                      = {10.1103/PhysRevD.87.094009},
  Eprint                   = {1212.6164},
  Primaryclass             = {hep-ph},
  Reportnumber             = {DESY-12-166, OUTP-12-19P},
  Slaccitation             = {%%CITATION = ARXIV:1212.6164;%%}
}

@Article{Hautmann:2017fcj,
  Title                    = {{Collinear and TMD Quark and Gluon Densities from Parton Branching Solution of QCD Evolution Equations}},
  Author                   = {Hautmann, F. and Jung, H. and Lelek, A. and Radescu, V. and Zlebcik, R.},
  Journal                  = {JHEP},
  Year                     = {2018},
  Pages                    = {070},
  Volume                   = {01},

  Archiveprefix            = {arXiv},
  Doi                      = {10.1007/JHEP01(2018)070},
  Eprint                   = {1708.03279},
  Primaryclass             = {hep-ph},
  Reportnumber             = {DESY-17-118},
  Slaccitation             = {%%CITATION = ARXIV:1708.03279;%%}
}

@Article{Hautmann:2017xtx,
  Title                    = {{Soft-gluon resolution scale in QCD evolution equations}},
  Author                   = {Hautmann, F. and Jung, H. and Lelek, A. and Radescu, V. and Zlebcik, R.},
  Journal                  = {Phys. Lett.},
  Year                     = {2017},
  Pages                    = {446-451},
  Volume                   = {B772},

  Archiveprefix            = {arXiv},
  Doi                      = {10.1016/j.physletb.2017.07.005},
  Eprint                   = {1704.01757},
  Primaryclass             = {hep-ph},
  Reportnumber             = {DESY-16-174},
  Slaccitation             = {%%CITATION = ARXIV:1704.01757;%%}
}

@Article{Marchesini:1987cf,
  Title                    = {{Monte Carlo Simulation of General Hard Processes with Coherent QCD Radiation}},
  Author                   = {Marchesini, G. and Webber, B. R.},
  Journal                  = {Nucl. Phys.},
  Year                     = {1988},
  Pages                    = {461-526},
  Volume                   = {B310},

  Doi                      = {10.1016/0550-3213(88)90089-2},
  Reportnumber             = {CAVENDISH-HEP-87-8, UPRF-87-212},
  Slaccitation             = {%%CITATION = NUPHA,B310,461;%%}
}

@article{Webber:1986mc,
    author = "Webber, B. R.",
    title = "{Monte Carlo Simulation of Hard Hadronic Processes}",
    reportNumber = "HEP-86-3",
    doi = "10.1146/annurev.ns.36.120186.001345",
    journal = "Ann. Rev. Nucl. Part. Sci.",
    volume = "36",
    pages = "253--286",
    year = "1986"
}

@article{Abdulov:2021ivr,
    author = "Abdulov, N. A. and others",
    title = "{TMDlib2 and TMDplotter: a platform for 3D hadron structure studies}",
    eprint = "2103.09741",
    archivePrefix = "arXiv",
    primaryClass = "hep-ph",
    reportNumber = "DESY 21-026, DESY-21-026, IFJPAN-IV-2021-4, JLAB-THY-21-3337",
    doi = "10.1140/epjc/s10052-021-09508-8",
    journal = "Eur. Phys. J. C",
    volume = "81",
    number = "8",
    pages = "752",
    year = "2021"
}

@Article{Hautmann:2014kza,
  author        = {Hautmann, F. and Jung, H. and Kraemer, M. and Mulders, P. J. and Nocera, E. R. and Rogers, T. C. and Signori, A.},
  title         = {{TMDlib and TMDplotter: library and plotting tools for transverse-momentum-dependent parton distributions}},
  journal       = {Eur. Phys. J.},
  year          = {2014},
  volume        = {C74},
  pages         = {3220},
  archiveprefix = {arXiv},
  doi           = {10.1140/epjc/s10052-014-3220-9},
  eprint        = {1408.3015},
  primaryclass  = {hep-ph},
  reportnumber  = {DESY-14-059, NIKHEF-2014-024, YITP-SB-14-24},
  slaccitation  = {%%CITATION = ARXIV:1408.3015;%%},
}

@article{Scimemi:2018xaf,
    author = "Scimemi, Ignazio and Vladimirov, Alexey",
    title = "{Systematic analysis of double-scale evolution}",
    eprint = "1803.11089",
    archivePrefix = "arXiv",
    primaryClass = "hep-ph",
    doi = "10.1007/JHEP08(2018)003",
    journal = "JHEP",
    volume = "08",
    pages = "003",
    year = "2018"
}

@Article{CASCADE:2021bxe,
  author        = {Baranov, S. and others},
  title         = {{CASCADE3 A Monte Carlo event generator based on TMDs}},
  journal       = {Eur. Phys. J. C},
  year          = {2021},
  volume        = {81},
  number        = {5},
  pages         = {425},
  archiveprefix = {arXiv},
  collaboration = {CASCADE},
  doi           = {10.1140/epjc/s10052-021-09203-8},
  eprint        = {2101.10221},
  primaryclass  = {hep-ph},
  reportnumber  = {DESY-21-005},
}

@Article{BermudezMartinez:2018fsv,
  author        = {Bermudez Martinez, A. and others},
  title         = {{Collinear and TMD parton densities from fits to precision DIS measurements in the parton branching method}},
  journal       = {Phys. Rev. D},
  year          = {2019},
  volume        = {99},
  number        = {7},
  pages         = {074008},
  archiveprefix = {arXiv},
  doi           = {10.1103/PhysRevD.99.074008},
  eprint        = {1804.11152},
  primaryclass  = {hep-ph},
  reportnumber  = {DESY 18-042, DESY-18-042},
}

@article{Hautmann:2007uw,
    author = "Hautmann, F.",
    title = "{Endpoint singularities in unintegrated parton distributions}",
    eprint = "hep-ph/0702196",
    archivePrefix = "arXiv",
    doi = "10.1016/j.physletb.2007.08.081",
    journal = "Phys. Lett. B",
    volume = "655",
    pages = "26--31",
    year = "2007"
}

@Article{Yang:2022qgk,
  author        = {Yang, H. and others},
  title         = {{Back-to-back azimuthal correlations in $\mathrm {Z} +$jet events at high transverse momentum in the TMD parton branching method at next-to-leading order}},
  journal       = {Eur. Phys. J. C},
  year          = {2022},
  volume        = {82},
  number        = {8},
  pages         = {755},
  archiveprefix = {arXiv},
  doi           = {10.1140/epjc/s10052-022-10715-0},
  eprint        = {2204.01528},
  primaryclass  = {hep-ph},
  reportnumber  = {DESY-22-025, CERN-TH-2022-113},
}

@article{Abdulhamid:2021xtt,
    author = "Abdulhamid, M. I. and others",
    title = "{Azimuthal correlations of high transverse momentum jets at next-to-leading order in the parton branching method}",
    eprint = "2112.10465",
    archivePrefix = "arXiv",
    primaryClass = "hep-ph",
    reportNumber = "DESY-21-219, LU-TP 21-53",
    doi = "10.1140/epjc/s10052-022-09997-1",
    journal = "Eur. Phys. J. C",
    volume = "82",
    number = "1",
    pages = "36",
    year = "2022"
}

@inproceedings{BermudezMartinez:2022tql,
    author = "Bermudez Martinez, A. and Hautmann, F.",
    title = "{Azimuthal di-jet correlations with parton branching TMD distributions}",
    booktitle = "{29th International Workshop on Deep-Inelastic Scattering and Related Subjects}",
    eprint = "2208.08446",
    archivePrefix = "arXiv",
    primaryClass = "hep-ph",
    reportNumber = "CERN-TH-2022-137, DESY-22-135",
    month = "8",
    year = "2022"
}

@inproceedings{BermudezMartinez:2021zlg,
    author = "Bermudez Martinez, A. and Hautmann, F. and Mangano, M. L.",
    title = "{Multi-jet physics at high-energy colliders and TMD parton evolution}",
    eprint = "2109.08173",
    archivePrefix = "arXiv",
    primaryClass = "hep-ph",
    month = "9",
    year = "2021"
}

@inproceedings{xFitter:2022zjb,
    author = "Abdolmaleki, H. and others",
    collaboration = "xFitter",
    title = "{xFitter: An Open Source QCD Analysis Framework. A resource and reference document for the Snowmass study}",
    eprint = "2206.12465",
    archivePrefix = "arXiv",
    primaryClass = "hep-ph",
    month = "6",
    year = "2022"
}

@article{Drell:1970wh,
    author = "Drell, S. D. and Yan, Tung-Mow",
    title = "{Massive Lepton Pair Production in Hadron-Hadron Collisions at High-Energies}",
    reportNumber = "SLAC-PUB-0755",
    doi = "10.1103/PhysRevLett.25.316",
    journal = "Phys. Rev. Lett.",
    volume = "25",
    pages = "316--320",
    year = "1970",
    note = "[Erratum: Phys.Rev.Lett. 25, 902 (1970)]"
}

@article{ATLAS:2023lhg,
    author = "Aad, Georges and others",
    collaboration = "ATLAS",
    title = "{A precise determination of the strong-coupling constant from the recoil of $Z$ bosons with the ATLAS experiment at $\sqrt{s} = 8$ TeV}",
    eprint = "2309.12986",
    archivePrefix = "arXiv",
    primaryClass = "hep-ex",
    month = "9",
    year = "2023"
}

@inproceedings{Raicevic:2024obe,
    author = "Raicevic, Natasa",
    title = "{Non-Perturbative Contributions to Low Transverse Momentum Drell-Yan Pair Production Using the Parton Branching Method}",
    booktitle = "{13th International Conference on New Frontiers in Physics}",
    eprint = "2412.00892",
    archivePrefix = "arXiv",
    primaryClass = "hep-ph",
    month = "12",
    year = "2024"
}

@article{Billis:2024dqq,
    author = "Billis, Georgios and Michel, Johannes K. L. and Tackmann, Frank J.",
    title = "{Drell-Yan Transverse-Momentum Spectra at N$^3$LL$'$ and Approximate N$^4$LL with SCETlib}",
    eprint = "2411.16004",
    archivePrefix = "arXiv",
    primaryClass = "hep-ph",
    reportNumber = "DESY-23-081, MIT-CTP 5572, Nikhef 2024-007",
    month = "11",
    year = "2024"
}
\end{document}